\documentclass[12pt]{article}
\usepackage{newinutile}
\usepackage{amsmath,amssymb,amsthm,amsfonts}
\usepackage{comment}
\usepackage{epsfig,graphics,color,calc,graphicx}
\usepackage{subfigure}
\usepackage{pgf,tikz}
\usepackage{mathrsfs}
\usepackage{color}
\usetikzlibrary{arrows}
\usepackage{mathrsfs}
\usepackage{hyperref}
\hypersetup{colorlinks=true,linkcolor=blue}
\usepackage{cite}
\usepackage{mathrsfs}

\newlength{\pecettawidth}
\begin{document}
\title{Conditional expectation of the duration of the classical 
gambler problem with defects}

\author{Alessandro Ciallella}
\email{alessandro.ciallella@uniroma1.it}
\affiliation{Dipartimento di Scienze di Base e Applicate per l'Ingegneria, 
             Sapienza Universit\`a di Roma, 
             via A.\ Scarpa 16, I--00161, Roma, Italy.}

\author{Emilio N.M.\ Cirillo}
\email{emilio.cirillo@uniroma1.it}
\affiliation{Dipartimento di Scienze di Base e Applicate per l'Ingegneria, 
             Sapienza Universit\`a di Roma, 
             via A.\ Scarpa 16, I--00161, Roma, Italy.}


\begin{abstract}
The effect of space inhomogeneities on a diffusing particle is studied 
in the framework of the 1D random walk.
The typical time needed by a particle to cross a one--dimensional 
finite lane, the so--called residence time,
is computed possibly in presence of a drift.
A local inhomogeneity is introduced as a single defect site 
with jumping probabilities differing from those at all the 
other regular sites of the system. 
We find complex behaviors in the sense that the residence time 
is not monotonic as a function of some parameters of the 
model, such as the position of the defect site. 
In particular we show that introducing at suitable 
positions a defect opposing to the 
motion of the particles decreases the residence time, i.e., 
favors the flow of faster particles. 
The problem we study in this paper is strictly connected to the 
classical gambler's ruin problem, indeed, it can be thought as that 
problem in which the rules of the game are changed when the gambler's fortune 
reaches a particular a priori fixed value.
The problem is approached both numerically, via 
Monte Carlo simulations, and analytically with two 
different techniques yielding different representations of the exact result. 
\end{abstract}



\ams:{82B41}

\maketitle

\section{Introduction} 
\label{ENMC:sec:introduction}
\par\noindent
Particles flowing across obstacles exhibit
interesting effects 
\cite{CPamm2017}. 
The dynamics, depending on the details of the systems, can be 
either accelerated or slowed down.
Anomalous diffusion,
i.e.,
sub--linear behavior with respect to time of the mean square distance 
traveled by particles undergoing Brownian motion,
is observed in cells and can be explained as an effect 
of macromolecules 
playing the role of obstacles for the diffusing smaller molecules
\cite{Sbj1994,HFrpp2013,MHSbj2017,ESMCBjcp2014}.
In many other different contexts it has been observed
that obstacles 
surprisingly accelerate the dynamics.
An obstacle placed close to the exit, 
can improve the 
out--coming flow in a granular system in presence 
of clogging \cite{TLPprl2001,ZGMPPpre2005,AMAGTOKpre2012,ZJGLAMprl2011}. 
Similar phenomena are observed in pedestrian flows
\cite{Hrmp2001,HFMV2011,CMpA2013,
MCKBud2014,CMcrm2012,CCCMm3as2018}
in case of panic,
where clogging at the door can be reduced by means of suitably 
positioned obstacles \cite{ABCKsjap2016,HFVn2000,HBJWts2005,EDLR2003}.

These phenomena are discussed here in the very basic scenario of the 
1D  symmetric random walk in connection with the behavior, 
as a function of the properties of the obstacle, 
of the typical time that a particle 
started at the left end of the lane 
needs to cross the whole lane.
This time will be called \emph{residence time}.
Here, the obstacle is represented by a single \emph{defect site}
characterized by different jumping probabilities with respect 
to those associated with all the other \emph{regular} sites.
This 1D model can be also interpreted as
a toy model for the 2D finite strip in which a rectangular obstacle 
is placed inside the strip in such a way that its 
right--hand boundary is in touch with the right boundary of the strip. 
The defect site mimics the sites in 
the first column of the 2D strip on the left of the obstacle, indeed, 
the 2D walker in such a column has a probability to move to the right
smaller than the probability to move to the left. 
The sites on the right of the defect are regular, since when the 2D walker 
enters one of the two channels flanking the obstacle its probability 
to move to the right or to the left is no more influenced by the obstacle.

It is worth mentioning that the problem we study here shares some 
features with the so--called \emph{blockage problem}
\cite{CCMpre2016,JanLeb94,SLM15},
where one considers a 1D dynamics with a slow down bond or site. 
The main problem, 
there, is that of understanding the effect of the local slow down on the 
stationary current in the thermodynamics limit. 

The residence time issue, as described above, has been firstly raised in 
\cite{CKMSpre2016, CKMSSpA2016}, where the flow of particles 
in an horizontal strip
undergoing a random walk with exclusion rule 
has been considered \cite{FPvSpre2017}. 
One of the most interesting results investigated in those papers 
is the possibility to spot complex behaviors, in the sense 
that the residence time unexpectedly shows up to be a not monotonic 
function of some parameters of the obstacles. 
The phenomenon was interpreted there as a consequence of the hard core 
particle interaction that gives rise to peculiar stationary 
particle density profiles. 
On the other hand, 
the studies reported in \cite{CCSpre2018,CCkrm2018}
show that, even for not interacting particles,
complex residence time behaviors can be observed as purely 
geometric effects. 
In \cite{CCkrm2018} the framework of Kinetic Theory was adopted
and a model with particles moving according to the linear 
Boltzmann dynamics was studied. 
In \cite{CCSpre2018} the simple symmetric random walk was considered
both in one and two dimensions. The one--dimensional 
reduction of the 2D obstacle problem was constructed 
by considering a 1D simple symmetric random walk 
with two defect sites modelling the effect of the obstacle on the 
motion of the particles. 

In this paper
we consider a single defect, but we relax the symmetry 
assumption and consider driven random walks as well. 
The residence time problem we study in this paper 
can be rephrased in the language of 
the classical gambler's ruin problem. In such a classical problem
a gambler starts with initial fortune $x_0$ and at each time he plays 
his fortune is either incremented by one with probability $p$ or
decreased by one with probability $q=1-p$. Fixed the goal
fortune $L$, the gambler goes on playing till either his 
fortune becomes $L$ or he goes bankrupt, namely his fortune 
becomes zero. We say that the gambler wins if his fortune 
becomes $L$ before going bankrupt, otherwise we say that the gambler loses.
The probability 
that the gambler reaches his target $L$ before going bankrupt, 
in the random walk language, 
is the probability that the particle started at site $x_0$ 
reaches the site $L$ before visiting the site zero.  
In the random walk language the residence time is the 
mean time needed by a particle started at site one to reach 
site $L$ conditioned to the fact that site $L$ is reached 
before the particle visits site zero. 
This quantity, in the gambler's language, is the 
mean duration of the game for the gambler with initial fortune $1$
conditioned to the fact that the gambler wins. 
In the sequel we shall define the model as a random walk 
model, but we shall often use its gambler's interpretation.

A classical reference for the gambler's ruin problem is the 
book \cite{Fbook1968} where the winning probability is 
studied. In \cite{S1975} the author investigates 
the conditioned duration (residence time).
In \cite{L2009} 
both the winning probability and the conditioned duration of the 
game are computed 
in the case $0<p+q<1$, that is to say, considering the 
possibility of ties. 
Here, in the most general case, 
we address and solve the winning probability and the 
conditioned duration of the game questions introducing 
a defect site, that is to say, in the gambler's language, 
the winning and loosing probabilities in a particular round 
differ from those characterizing all the other rounds. 
We perform both a numerical and an exact computation of the 
winning probability and of the conditioned duration of the 
game. The exact computation is performed both using 
the generating function method proposed in \cite{Fbook1968,S1975,L2009}
and the method developed in \cite{CCSpre2018} to attack the problem 
with two defects.

Using the random walk language, 
we discuss the dependence of the residence time on the defect 
parameters both in the symmetric and asymmetric case, $p=q$ and $p\neq q$ respectively. 
In particular we find a complex behavior of the residence time 
with respect to the position of the defect, namely, we 
show that the residence time can be either increased or decreased, 
with respect to the no defect case, depending on the position 
of the obstacle. This effect is found both in the symmetric and 
in the asymmetric cases, but its magnitude is more important 
in the first case. 
In the gambler language, this means that the mean duration of the 
game, conditioned to the fact that the players wins, changes
if the rounds played when the gambler's fortune has an a priori fixed  
value are played with different rules. This is obvious, 
but highly not trivial is the fact that depending on the value of the 
fortune at which the rules are 
modified the duration of the game can either increase or decrease. 
We also stress that the dependence of the residence time on the 
position of the defect is not symmetric when the defect is moved along the 
lane, as it was remarked in \cite{CCSpre2018} this could be connected 
with the possibility to observe uphill currents 
\cite{CCpre2017,ACCGprep2018}
in presence of 
obstacles.

The paper is organized as follows. 
In Section~\ref{s:1D} we define the model. In Section~\ref{s:dimo} 
we find two exact representations of the residence time. 
In Section~\ref{s:discussione} we compare our exact results to 
the Monte Carlo estimate of the residence time and we discuss 
our findings. 

\section{The model}
\label{s:1D} 
\par
\begin{figure}[t]
\vskip 0.3 cm
\centerline{%
{\includegraphics[width=.4\textwidth]{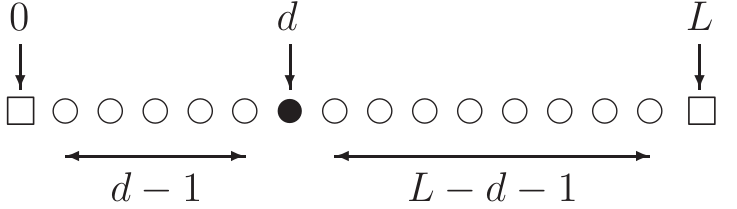}}
}
\caption{Representation of the lattice on which the model 
is defined. Open and solid circles
respectively represent regular and singular (defect) sites. 
The two squared box represent the two boundary 
absorbing sites. 
}
\label{f:1}
\end{figure}

\noindent
We consider a simple random walk on $\{0,1,\dots,L\}$, i.e.,
the set $[0,L]\cap\mathbb{Z}$.
The sites $0$ and $L$ are absorbing, 
so that when the particle reaches one of these two sites the walk is stopped. 
All the sites $1,\dots,L-1$ are \emph{regular} except for the 
site $d$ called \emph{defect} or \emph{singular} site
(see Figure~\ref{f:1}), with 
$d\in\{2,\dots,L-2\}$.
The parameter $d$ is chosen in such a way that the 
defect site cannot be $1$ or $L-1$.
The number of regular sites on the left (resp.\ right) 
of the defect site is $d-1$ (resp.\ $L-d-1$).

At each unit of time the walker 
jumps to a neighboring site according to the following rule:
if it is on a regular site, then it 
performs a simple 
random walk with probabilities $p$ to right, $q$ to left. 
If $p+q <1$ the walker does not move with probability $1-p-q$.
When it is at the defect site it jumps with 
probability $p'$ to the right, with probability $q'$ 
to the left, and with probability $\varepsilon$ it does not move.  
Here, $p'\in(0,1)$ and $\varepsilon\in[0,1)$ such that $p'+q'+\varepsilon=1$.
If $p+q=1$, $p'=p$, and $q'=q$ the classical \emph{gambler ruin} problem 
is recovered \cite{Fbook1968,S1975,L2009}.

The array $1,\dots,L-1$ will 
be called the \emph{lane}. The sites $0$ and $L$ will be, 
respectively, called the \emph{left} and \emph{right exit} of the lane. 

This 1D model can be also interpreted as
a toy model for the 2D finite strip in which a rectangular obstacle 
of width $L-d-1$ is placed inside the strip in such a way that its 
right--hand boundary is in touch with the right boundary of the strip. 
The defect site $d$ mimics the sites in 
the first column of the 2D strip on the left of the obstacle. Indeed, 
the 2D walker in such a column has a probability to move to the right
smaller since it could hit the obstacle. 
Let us stress that 
the sites $d+1,\dots,L-1$ are regular, since when the 2D walker 
enters one of the two channels flanking the obstacle its probability 
to move to the right or to the left is not influenced by the obstacle anymore.
The symmetric situation of a 2D obstacle of width $d-1$ and left--hand boundary in touch with the left boundary of the strip can be considered. In this case the decreasing  probability is the one to jump to the left from the column on the right of the obstacle.

The main quantity of interest is the residence time, which is 
defined by starting 
the walk at site $1$ and computing 
the typical time that the particle takes to reach the site $L$ conditioned 
to the fact that it 
reaches $L$ before $0$.
We let $x_t$ be the position of the walker at time $t$ and 
denote by $\mathbb{P}_k$ and $\mathbb{E}_k$ the probability associated to the 
trajectories of the walk and the related average operator 
for the walk started at $x_0=k$ with $k=1,\dots,L-1$. 
We let $T_i$ 
be the \emph{first hitting time to} $i$, namely, 
the first time at which the trajectory of the walk reaches the site $i$, 
with the convention that 
$T_i=\infty$ if the trajectory does not reach the site $i$. 
The main quantity of interest is the \emph{residence time} or
\emph{total residence time} 
\begin{equation}
\label{one010}
R
=
\mathbb{E}_1[T_L|T_L<T_0]
=
\sum_{t=1}^\infty
t\mathbb{P}_1[T_L=t|T_L<T_0]
\;\;.
\end{equation}
Note that the residence time is defined for the walk started at $x_0=1$ and 
the average is computed conditioning to the event $T_L<T_0$,  namely,
conditioning to the fact that the particle exits the lane through the 
right exit. 

\section{Exact computation of the residence time}
\label{s:dimo} 
\par\noindent
In this section we compute the residence time \eqref{one010} 
for the model introduced 
in Section~\ref{s:1D}.
We first deal with the crossing probability and then with the residence 
time. We use standard techniques developed in \cite{Fbook1968,S1975,L2009}
and compare the results with those obtained by applying the methods
developed in \cite{CCSpre2018} to tackle a similar model with two 
defects. 

\subsection{Crossing probability}
\label{s:cross} 
\par\noindent
For the classical gambler's ruin problem (no defect),
the probability of the gambler's ultimate ruin with initial fortune $i$, i.e.,
$t_i=\mathbb{P}_i[T_0<T_L]$, is given by 
\begin{displaymath}
t_i=\frac{(q/p)^L - (q/p)^i}{(q/p)^L-1} 
\;\textrm{ for }\;
p\neq q
\;\textrm{ and }\;
t_i= 1- (i/L) 
\;\textrm{ for }\;
p=q,
\end{displaymath}
see, for instance, \cite{Fbook1968}. In the random walk 
language $t_i$ is the 
probability that the walker started at site $i$ reaches the site 
$0$ before $L$. 

Here, 
we generalize this result to the problem stated in Section~\ref{s:1D}, 
namely, 
for the random walk on the lane with a defect site. 
Starting with an initial fortune $i$, after one trial the gambler 
fortune is $i+1$ or $i-1$, or $i$ if it doesn't move. 
It follows that
\begin{equation}\label{eq:differenze}
\begin{split}
&t_i= p t_{i+1}+ q t_{i-1} + (1-p-q)t_i,\quad \forall i=1,\ldots ,d-1,
\\ & t_d= p' t_{d+1} + q' t_{d-1} + (1-p'-q') t_d,
\\ & t_{j}= p t_{j+1} +q t_{j-1} + (1-p-q)t_j,\quad \forall j= d+1,\ldots ,L-1.
\end{split}
\end{equation}
Note that since the random walk ends in $0$ or $L$ respectively with the win or the loss of the gambler, $t_0=1$ and $t_L=0$.

To solve \eqref{eq:differenze}, we first note that 
the first set of equations \eqref{eq:differenze}
can be straightforwardly rewritten as
\begin{displaymath}
p t_i- qt_{i-1}= pt_{i+1} - q t_i
\;\textrm{ for }\;
i=1,\dots,d-1.
\end{displaymath}
Hence, 
we let $\sigma=p t_i- qt_{i-1}$ for $i=1,\ldots,d$, 
recall $t_0=1$, and, dividing by $p$, we get 
\begin{equation}
\label{eq:t_i}
\begin{split}
&t_1= \frac{\sigma}{p} + \frac{q}{p},
\\ 
& t_2= \frac{q}{p}t_1 + \frac{\sigma}{p}
= 
\Big(\frac{q}{p}\Big)^2 + \frac{q}{p}\frac{\sigma}{p} + \frac{\sigma}{p},
\\& \qquad \vdots
\\& t_d= \frac{q}{p}t_{d-1} 
+ 
\frac{\sigma}{p}
=
\Big(\frac{q}{p}\Big)^d 
+ 
\frac{\sigma}{p}\sum_{k=0}^{d-1} \Big(\frac{q}{p}\Big)^k.
\end{split}
\end{equation}
By treating similarly the second equation in \eqref{eq:differenze}, 
we find $(p'+q')t_d=p' t_{d+1} +q' t_{d-1}$, so that
\begin{equation}
\label{eq:t_s+1}
\begin{split}
t_{d+1}&= -(q'/p') t_{d-1} + ((p'+q')/p') t_{d}
= (q'/p') (t_d-t_{d-1}) + t_d 
\\ 
&= 
\frac{q'}{p'} \Big(\frac{q}{p}\Big)^{d-1} 
\Big(\frac{q}{p} - 1 + \frac{ \sigma}{p}\Big)
+\Big(\frac{q}{p}\Big)^d 
+\frac{\sigma }{p} \sum_{k=0}^{d-1}
\Big(\frac{q}{p}\Big)^k.
\end{split}
\end{equation}
The third family of equations \eqref{eq:differenze} allows us to write
\begin{equation}\label{t_j}
p t_{j+1} - q t_j = p t_j - q t_{j-1} =\lambda 
\;\textrm{ for }\; 
j=d+1,\ldots,L-1.
\end{equation}
Using the expressions 
\eqref{eq:t_i} and \eqref{eq:t_s+1} 
of $t_{d+1}$ and $t_d$ we can compute 
the explicit expression.
Then, we find the probabilities
\begin{equation}
\label{eq:t_j_explicit}
\begin{split}
&t_{d+2}= \frac{\lambda}{p} + \frac{q}{p} t_{d+1}
\\& 
t_{d+3}= \frac{\lambda}{p} + \frac{q}{p} t_{d+2}=
\Big(\frac{q}{p}\Big)^2 t_{d+1} 
+\frac{\lambda}{p}\Big(1+\frac{q}{p}\Big)
\\ 
&  \qquad \vdots
\\
& 
t_{L}=\Big(\frac{q}{p}\Big)^{L-d-1} t_{d+1} 
+ 
\frac{\lambda}{p} \sum_{k=0}^{L-d-2} \Big(\frac{q}{p}\Big)^k.
\end{split} 
\end{equation}
Finally, using $t_L=0$ we can find $\sigma$. Indeed, by letting 
$t_L=0$ and $\lambda =pt_{d+1} - qt_d$ in the last of equations 
\eqref{eq:t_j_explicit}, after some algebra,
we get
\begin{equation}
\label{eq:metto_t_L=0}
t_{d+1}\Big[\sum_{k=0}^{L-d-1} \Big(\frac{q}{p}\Big)^k\Big] 
- \frac{q}{p} t_d \Big[\sum_{k=0}^{L-d-2} \Big(\frac{q}{p}\Big)^k\Big]=0.
\end{equation}
To complete the computation we have to distinguish the two 
cases $p\neq q$ and $p=q$. The second one is easier and will be treated 
later. 

\par\noindent
\textit{Case $p\neq q$.\/} 
Equation \eqref{eq:metto_t_L=0} can be re written as
\begin{equation*}
\frac{1-(q/p)^{L-d}}{1-q/p} t_{d+1} - \frac{q/p-(q/p)^{L-d}}{1-q/p} t_d=0.
\end{equation*}
Thus,
multiplying by $1-(q/p)$ and using the explicit expressions of 
$t_{d+1}$ and $t_{d}$, after some algebra we get the expression 
of $\sigma$:
\begin{equation}\label{eq:sigma_expl}
\begin{split}
{\sigma}&=
p 
\Big\{ \Big[ \Big(\frac{q}{p}\Big)^{L-d}-1 \Big] 
\Big[ \frac{q'}{p'} \Big(\frac{q}{p}\Big)^{d-1} \Big( \frac{q}{p} - 1 \Big)   
+ 
\Big(\frac{q}{p}\Big)^d \Big] 
+ 
\Big(\frac{q}{p}\Big)^{d+1} 
-
\Big(\frac{q}{p}\Big)^L \Big\}
\\&
\phantom{=}
\times
\Big\{ 
\Big[1-\Big(\frac{q}{p}\Big)^{L-d}\Big] 
\Big[\frac{q'}{p'} \Big(\frac{q}{p}\Big)^{d-1} 
+ \frac{1-(q/p)^d }{1-q/p)}\Big]  
+ \Big[\Big(\frac{q}{p}\Big)^{L-d}
- \frac{q}{p}\Big]
\frac{1-(q/p)^d }{1-q/p} 
\Big\}^{-1}
\end{split}
\end{equation}

Once $\sigma$ is known, we can write explicitly each $t_i$. In particular 
we can write the probability $t_1=\mathbb{P}_1[T_0<T_L]$
that the particle started at $1$ exits the lane in $0$ 
and 
the probability $1-t_1=1-\mathbb{P}_1[T_0<T_L]$
that the particle started at $1$ exits the lane in $L$. 
Recalling that $t_1= \sigma/p+q/p$, 
some painful algebra yields  
\begin{equation}
\label{eq:finale}
1-t_1 
= 
\frac{p' (p-q)q}{p(-p'q(-1 + (q/p)^d)+ pq'(-(q/p)^L + (q/p)^d))}.
\end{equation}

Note  that the result does not change if at 
the regular sites the probabilities to jump right 
$p_i$, to jump left $q_i$ or to stay $r_i$ are not constant anymore, 
but the ratio $p_i/q_i = p/q$ is conserved. 
Indeed, the equation on a generic site becomes
\begin{displaymath}
t_i=(p_i+q_i+r_i) t_i =  p_i t_{i+1} + q_i t_{i-1} 
+ r_i t_i
\end{displaymath}
so that \eqref{eq:t_i}--\eqref{eq:sigma_expl} are still valid.

We remark that the probability $1-t_1$
can be also calculated by means of the Markov property 
following the ideas developed in \cite{CCSpre2018}. 
Let us define
\begin{displaymath}
p_1=\mathbb{P}_1(T_d<T_0),\;
p_2=\mathbb{P}_d(T_{d+1}<T_0),\;
\textrm{ and }\;
p_3=\mathbb{P}_{d+1}(T_L<T_0).
\end{displaymath}
By using the classical gambler's ruin problem results we find 
$p_1=({1 - (q/p)})({1-(q/p)^d})^{-1}$.
We observe that a random walk starting from the site $1$ reaches again 
the site $d$ before hitting $d+1$ for the first time with probability 
$(1-p'-q') +(q'(1-(q/p)^{d-1})(1-(q/p)^d)^{-1} ) $, where the second 
term in the sum is due to trajectories that reach $d-1$ after one step 
and than eventually again $d$ with probability 
$(1-(q/p)^{d-1})(1-(q/p)^d)^{-1}$.
So the probability $p_2$ reads
\begin{displaymath}
p_2= \sum_{k=0}^{+\infty}p'[(1-p'-q') +(q'(1-(q/p)^{d-1})(1-(q/p)^d)^{-1} )]^k
=\frac{p'}{p'+q' -q'\frac{1-(q/p)^{d-1}}{1-(q/p)^d} }.
\end{displaymath}
Finally, 
the probability $p_3$ can be expressed by means of the 
classical result on the gambler's ruin probability and the previous $p_2$.
Starting from $d+1$ the probability to reach $L$ before $d$  
is indeed $(1-(q/p))/(1-(q/p)^{L-d})$, while with 
probability $((q/p)^{L-d}-(q/p))/((q/p)^{L-d}-1)$ the walk 
reaches $d$ before $L$. The walk can return several times on site $d+1$, giving a geometric sum as in the calculation of $p_2$. So, after the summation we find
\begin{displaymath}
p_3= 
\frac{1-(q/p)}{1-(q/p)^{L-d}}
\frac{1}{1- p_2 \frac{(q/p)^{L-d}-(q/p)}{(q/p)^{L-d}-1}}
= \frac{(q/p)-1}{(q/p)^{L-d} -1 - [(q/p)^{L-d} - q/p]p_2}.
\end{displaymath}
Computing the product $p_1p_2p_3$, after some algebra, one gets the 
same expression $1-t_1$ given in \eqref{eq:finale}.

\par\noindent
\textit{Case $p= q$.\/} 
In the symmetric case the expression \eqref{eq:metto_t_L=0} becomes
\begin{equation}\label{eq_sig_sym}
t_{d+1}+(t_{d+1}-t_d)(L-d-1)=0 
\end{equation}
where $t_{d+1}=q'/p'\cdot \sigma/p + 1 + d \sigma/p $ and $t_d=1 +d \sigma/p$.
Hence, we get
\begin{equation}
\sigma= -\frac{p}{d+ (L-d)q'/p'}.
\end{equation}
Thus, we find 
$t_1=1-1/(d+ (L-d)q'/p')$ and the probability to exit from 
the right side reduces to $1-t_1=1/(d+ (L-d)q'/p')$.
We can get the same result by computing 
$p_1$, $p_2$, and $p_3$ as above.
Here, $p_1=1/d$, $p_2=p'/(1-[q' (d-1)/d  + 1-q'-p'])$,
and $p_3=1/(1 + (L-d-1)(1-p_2))$.

\newpage 
\subsection{Residence time: the generating function approach}
\label{s:residence-gen} 
\par\noindent
We call $u_{n,i}$ the probability to exit the walk from the left side, starting from $i$ and after $n$ steps.  
We construct   the generating function of the probability of exiting from the 
left side starting from $i$,   $U_i(s)=\sum_{n\geq 0} u_{n,i} s^{n}$,
see \cite{Fbook1968,S1975,L2009}.
Note also that $t_i=U_i(1)$.
Therefore, the series defining $U_i(s)$ is totally and
thus uniformly convergent for $s\in[0,1]$.
Since the derivative of the generating function is 
$U_i'(s)=\sum_{n\geq1} n u_{n,i} s^{n-1}$, 
following 
\cite{S1975}
we have that 
\begin{equation}
\label{res000}
\lim_{s\to 1^-} U_i'(s)
= 
\sum_{n\geq1} n u_{n,i}
=
t_i E_i,
\end{equation}
where
$E_i $ is the conditional expectation of the duration of the game given that 
the random walk ends in $0$ and it is finite for any fixed 
$p$, $q$, $p'$, and $q'$.

Following the approach of \cite{Fbook1968, L2009}, we find the generating function as the solution of a system combining the equations for the generating function in the bulks (regular sites) and on the singular site.

Recalling that on the regular sites $p$ and $q$ are, respectively, 
the probabilities to jump to the right and to the left,
we find that 
\begin{equation}\label{eq:u_n,i}
u_{n+1,i}=p u_{n,i+1} + q u_{n,i-1} + (1-p-q) u_{n,i}
\end{equation}
on the regular sites, namely, for $i=1,\dots,d-1$ and 
$i=d+1,\dots,L-1$,
and boundary values 
\begin{equation*}
u_{n,0}=u_{n,L}=0 \,\, \textrm{when} \,\, n\geq 1,
\quad u_{0,0}=1,\quad u_{0,i}=0  \,\, \textrm{when} \,\, i\geq 1.
\end{equation*}
Thus, multiplying \eqref{eq:u_n,i} by $s^{n+1}$ and summing for 
$n=0,1,2,\ldots$ we find the following equation in the bulk, i.e., 
$i=1,\dots,d-1$ and $i=d+1,\dots,L-1$,
\begin{equation}
\label{eq:U_i_bulk}
U_i(s)= ps U_{i+1}(s) + qs U_{i-1}(s) + (1-p-q) s U_{i}(s),
\end{equation}
to be solved with boundary conditions
$U_0(s)=1$ and $U_L(s)=0$.
The equation for $U_d(s)$ on the defect site is 
\begin{equation}
\label{eq:U_defect}
U_d(s)= p's U_{d+1}(s) + q's U_{d-1}(s) + \varepsilon s U_{i}(s),
\end{equation}
since the probabilities on the singular site are $p'$ to jump to the right, 
$q'$ to the left, and $\varepsilon$ to stay.

It is known that in the bulk 
of regular sites the generating function $U_i$ for fixed $s$ can be 
searched in the form $\lambda^{i}(s)$, except for the case $p=q$ and $s=1$ at the same moment, where the generating function is linear, see later.
Substituting in \eqref{eq:U_i_bulk} it is found
\begin{equation}\label{tre010}
\lambda_\pm(s)
=
\frac{1-(1-p-q)s \pm \sqrt{[1-(1-p-q)s]^2 - 4pqs^2}}{2ps}
\end{equation}
The generating function in the bulk can be written as a linear combination 
of terms in the form $\lambda_-^i$ and $\lambda_+^i$  
for $i=0,\dots,L$.
We consider two different linear combinations in the bulk on the left and 
on the right of the defect site, namely, we introduce two coefficients 
on the left and two (possibly) different coefficients on the right.
Thus we find two different representation of $U_d(s)$.
Therefore, we will be able to find the unknown coefficients by 
requiring that these two representation are equal, 
that the equation at the defect \eqref{eq:U_defect}, and the boundary 
conditions are satisfied. 
More precisely,  the generating function reads
\begin{equation}\label{tre015}
\begin{split}
&U_i(s)= G(s) \lambda_+^i{(s)} + H(s) \lambda_-^i(s), \quad i=0,1,\ldots,d;
\\&U_j(s)= U(s) \lambda_+^i{(s)} + V(s) \lambda_-^i(s), \quad j=d,d+1,\ldots,L,
\end{split}
\end{equation}
and the coefficients $G(s)$, $H(s)$, $U(s)$, and $V(s)$ solve the system
\begin{equation}\label{tre020}
\begin{cases}
& G \lambda_+^d + H \lambda_-^d = U \lambda_+^d + V \lambda^d_-
\\ & (1-\varepsilon s ) (G \lambda_+^d + H \lambda_-^d)= q' s (G \lambda_+^{d-1} + H \lambda_-^{d-1} ) + p's (U \lambda_+^{d+1} + V \lambda^{d+1}_-)
\\& 1=G+H
\\& 0 = U\lambda_+^L + V\lambda_-^L.
\end{cases}
\end{equation}
The system \eqref{tre020} has a unique solution $(G,H,U,V)$ that can be explicitly expressed in terms of  jump probabilities, $\lambda_{\pm}$ and $s$.
Thus, by substituting $\lambda_{\pm}$ given by the \eqref{tre010} in the formulas \eqref{tre015}, we find the explicit expressions of the generating function $U_i(s)$. 
However, due to the length and complexity of these expressions, we prefer not to report here the solutions of the system in the general case.
Note that in the easiest case, i.e. if the site $d$ is regular, namely $p'=p$ and $q'=q$, 
the solutions are $G=U$ and $H=V$ in the form of the result in
\cite[above equation (2)]{L2009}, more precisely we find
\begin{displaymath}
G=U=
\frac{1}{1-\lambda_+^L\lambda_-^{-L}}
\;\;\textrm{ and }\;\;
H=V=
\frac{1}{1-\lambda_+^{-L}\lambda_-^L}
\;,
\end{displaymath}
which reduce to the classical ones in
Feller \cite[equation (4.10) in Paragraph~XIV.4]{Fbook1968} in the case
$p+q=1$.

The conditional expectation $E_i$ of the duration of the game starting 
from $i$ and ending in $0$ can be computed using 
equation \eqref{res000}.
The limit allows us to include in this formula even the symmetric 
case $p=q$, where the generating function has not the form of 
a combination of powers $\lambda^i$ anymore for $s=1$.

The conditional expectation $F_i$ of the duration of the game starting 
from $i$ and ending in $L$, in particular starting from $i=1$, can be 
now deduced from the $U_{L-1}$, by exchanging the role of $p$ and $q$, $p'$
and $q'$, $i$ and $L-i$. 
In particular $F_1$ will be the  representation of the residence time by means of the
generating function method.

\subsection{Residence time: the local times approach}
\label{s:residence-loc} 
\par
\begin{figure}[t]
\vspace{0.5 cm}
\centerline{%
\hspace{-0.5 cm}
{\includegraphics[width=.3\textwidth]{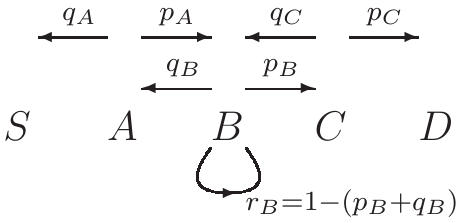}}
}
\caption{Schematic representation of the five state chain model.
}
\label{f:5stati}
\end{figure}

\noindent
A different approach to calculate the residence time is to decompose 
the expected duration of the game as a sum of local times spent on each site
and 
to make use of a reduced picture via a five states chain. 
This method has been proposed in 
\cite{CCSpre2018} to study a similar problem with two 
defect sites. We report it here in detail for the sake of completeness.
We first consider the five states, named 
$S$, $A$, $B$, $C$, and $D$,
chain 
with jump probabilities  
depicted in figure~\ref{f:5stati}. The chain is started at time 
$0$ in $B$. 
The probability $Q_k$, with $k\ge1$, for the 
chain to reach $D$ before $S$ and 
return exactly $k-1$ times  
to the site $B$ before reaching $D$ is 
\begin{equation}
\label{dimo000}
Q_k
=
p_Bp_C [r_B+q_Bp_A+p_Bq_C]^{k-1}
\;\;,
\end{equation}
where $r_B=1-(p_B+q_B)$.
Indeed, see \cite{CCSpre2018},
\begin{displaymath}
\begin{split}
Q_k
&
=
p_Bp_C 
\sum_{n=0}^{k-1} 
{{k-1}\choose{n}} (q_Bp_A)^{k-1-n}
\sum_{s=0}^n
{{n}\choose{s}} 
 (p_Bq_C)^s (r_B)^{n-s}
\end{split}
\end{displaymath}
where $s$ counts the number of times that, starting from $B$, 
the chain either jumps to $C$ or it stays in $B$ and 
$n$ counts the number of times that starting from $B$ it jumps to $A$. 
The equation \eqref{dimo000} is then proven by 
using the binomial theorem.

Now, 
to compute the duration of the game defined in Section~\ref{s:1D},
we focus on the time spent by a trajectory at any site $i$, namely, 
we consider the 
\emph{local times}
$\tau_i =|\{t>0:\, x_t=i\}|$
for any $i=1,\dots, L-1$, where 
$|\cdot|$ denotes the cardinality of a set. 
The fist arrival time to $L$, provided that it exists, 
can be expressed as the sum of local times on the sites reached by the walk.
Hence the conditional expectation $F_1$, i.e., the residence time, 
reads 
\begin{equation}
\label{dimo030}
F_1=
\sum_{i=1}^{L-1}\mathbb{E}_1[\tau_i|T_L<T_0]
\end{equation}
where $T_L<T_0$ is the event that the walk reaches $L$ before $0$, 
i.e., it exits from the left side.
As proven in \cite{CCSpre2018},
for all $i \in \{1,\ldots,L-1\}$
it holds
\begin{equation}
\label{dimo040}
\begin{split}
&\mathbb{E}_1[\tau_i|T_L<T_0]
=\frac{\mathbb{P}_1[T_i<T_0]}{\mathbb{P}_1[T_L<T_0]}
\frac{p_Bp_C}{[1-(r_B+p_Bq_C+q_Bp_A)]^2}
\;\;,
\end{split}
\end{equation}
where 
\begin{equation}
\label{dimo050}
\begin{array}{ll}
p_A=\mathbb{P}_{i-1}(T_i<T_0),
&
q_A=\mathbb{P}_{i-1}(T_0<T_i),\\
p_B=\mathbb{P}_i(x_1=i+1),
&
q_B=\mathbb{P}_i(x_1=i-1),\\
p_C=\mathbb{P}_{i+1}(T_L<T_i),
&
q_C=\mathbb{P}_{i+1}(T_i<T_L).
\end{array}
\end{equation}
Note that $p_A+q_A=1$, $p_{B} + q_{B} + r_{B} = 1$, and $p_C+q_C=1$.

Our strategy to compute the residence time is the following: 
for any $i=1,\dots,L-1$ we shall compute 
$\mathbb{E}_1[\tau_i|T_L<T_0]$ identifying the correct 
values of $p_A$, $q_A$, $p_B$, $q_B$, $p_C$, and $q_C$ to be used,
whose definition depends on the choice of the site $i$. Finally, 
the sum \eqref{dimo030} will provide us with the 
residence time. 

Note that the values of $p_A$,$	\ldots$,$q_C$ have to be searched in the 
form of $p_1$, $p_2$, or $p_3$ defined in Section~\ref{s:cross} depending 
on the value of $i$.
The site $i$ can be chosen in five different ways: as a site in the bulk 
before or beyond the defect, as a site neighboring the defect or as the 
defect site.

\emph{Case $i\leq d-2$.} The probability $\mathbb{P}_1[T_i<T_0]$ follows from the classic results on the gambler's ruin problem. It is $(1-q/p)/(1-(q/p)^i)$ if $p\neq q$ and $1/i$ if $p=q$.
Analogously,
$p_A$ follows from the same result but starting from $i-1$, 
so $p_A=(1-(q/p)^{i-1})/(1-(q/p)^i)$ if $p\neq q$ and $(i-1)/i$ if $p=q$. 
On the other hand, 
$q_A$ can be deduced as $q_A=1-p_A$, whereas $p_B$, $q_B$,  
and $r_B$ are respectively $p$, $q$, and $1-p-q$.
Finally, $p_C$ is the crossing probability for a shorter lane, with 
$L$ replaced by $L-i$ and initial site $i+1$. From \eqref{eq:finale} we 
thus get 
\begin{displaymath}
p_C=\frac{p' (p-q)q}{p(-p'q(-1 + (q/p)^{d-i})+ p (-(q/p)^{L-i} + (q/p)^{d-i})q')}
\end{displaymath}
for $p\neq q$, while if $p=q$ it holds $p_C=1/(d-i+ (L-d-i)q'/p')$. 
We also have $q_C=1-p_C$.

\emph{Case $i= d-1$.}
The probability $\mathbb{P}_1[T_{d-1}<T_0]$ doesn't change its form
and becomes $(1-q/p)/(1-(q/p)^{d-1})$ if $p\neq q$ and $1/(d-1)$ if $p=q$. 
In the same way we find $p_A=(1-(q/p)^{d-2})/(1-(q/p)^{d-1})$ if $p\neq q$ 
and $(d-2)/(d-1)$ if $p=q$. 
The terms 
$p_B$, $q_B$, and $r_B$ are the same as before, 
whereas $q_C$ has the structure of $p_2$ of Section~\ref{s:cross} 
but exchanging the role of $p$ and $q$, $p'$ and $q'$ and the sites 
before and beyond the defect.
Thus, if $p\neq q$
\begin{displaymath}
q_C=\frac{q'}{p'+q' -p'[1-(p/q)^{L-d-1}]/[1-(p/q)^{L-d}]}.
\end{displaymath}
In the symmetric case, $p=q$, $q_C$ reads $q'/(1-[p' (L-d)/(L-d+1)+1-q'-p'])$.

\emph{Case $i= d$.} The probability $\mathbb{P}_1[T_d<T_0]$ is 
$(1-q/p)/(1-(q/p)^{d})$ if $p\neq q$ and $1/d$ if $p=q$.
As in the previous cases, 
$p_A=(1-(q/p)^{d-1})/(1-(q/p)^{d})$ if $p\neq q$ and $(d-1)/(d)$ if $p=q$.
Now, $p_B=p'$, $q_B=q'$, and $r_B=\varepsilon$.
Finally, 
$p_C$ is easily find as a gambler's win probability, 
$(1-q/p)/(1-(q/p)^{L-d})$ or $1/(L-d)$ for $p\neq q$ and $p=q$ respectively.

\emph{Case $i=d+1$.} 
First note that 
$\mathbb{P}_1[T_{d+1}<T_0]$ is exactly the product $p_1 p_2$.
Moreover, 
$p_A=p_2$, $p_B=p$, $q_B=q$ and $r_B=1-p-q$.
Finally, 
$p_C$ is again as a gambler's win probability, 
$(1-q/p)/(1-(q/p)^{L-d})$ or $1/(L-d)$ for $p\neq q$ 
and $p=q$ respectively.

\emph{Case $i\geq d+2$.} 
First note that 
$\mathbb{P}_1[T_{i}<T_0]$ and $q_A$ have the form of a crossing probability 
for a shorter lane, namely, 
\begin{displaymath}
\mathbb{P}_1[T_{i}<T_0]= 
\frac{p' (p-q)q}{p(-p'q(-1 + (q/p)^d)+ p (-(q/p)^{i} + (q/p)^d)q')}
\end{displaymath}
when $p\neq q$ and it is equal to $1/(d+ (L-d-i)q'/p')$ if $p=q$.
To find $q_A$ we exchange the role of the probability to jump to the 
left and to the right, and of sites before and beyond the defect.
Thus 
\begin{displaymath}
q_A=\frac{q' (q-p)p}{q(-q'p(-1 + (p/q)^{i-d})+ q (-(p/q)^{i} + (p/q)^{i-d})p')}
\end{displaymath}
when $p\neq q$ and $1/(L-d-i+ i p'/q')$ if $p=q$.
The probabilities in $B$ and $C$ has the same form of the previous case $i=d+1$.

\section{Discussion}
\label{s:discussione} 
\par\noindent
In this section we investigate numerically the residence time 
problem and 
compare the numerical experiments results
to the 
exact computations carried out in 
Section~\ref{s:dimo}.
We compute numerically the residence time 
by simulating many particles and averaging 
the time that each of them takes to exit through the right 
ending point. 
The particles exiting through the left ending point are discarded.

We shall discuss our 
findings providing plots where 
the solid lines will represent the numerical result 
and the dots will denote the exact results obtained with the two strategies
discussed in Section~\ref{s:dimo}.
More precisely, the open and solid dots will denote
the representations of the 
residence time obtained
by using, respectively, the local time approach and the generating function method. 
All the details about the numerical simulations  
are in the figure captions. The statistical error, since negligible, 
is not reported in the picture. 
We fix from now on as the length of the lane $L=100$.

\begin{figure}[ht!]
\begin{picture}(200,290)(0,0)
\put(-50,145){
  \includegraphics[width=0.8\textwidth]{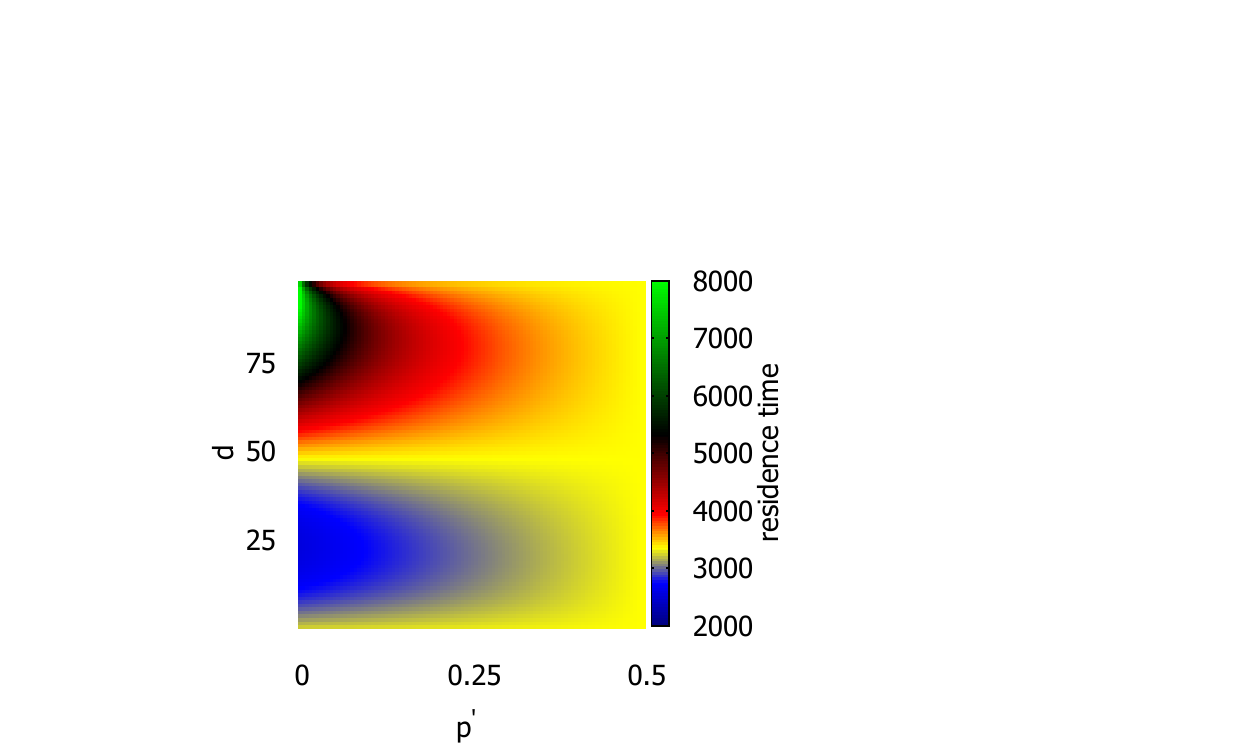}
}
\put(175,145){
  \includegraphics[width=0.8\textwidth]{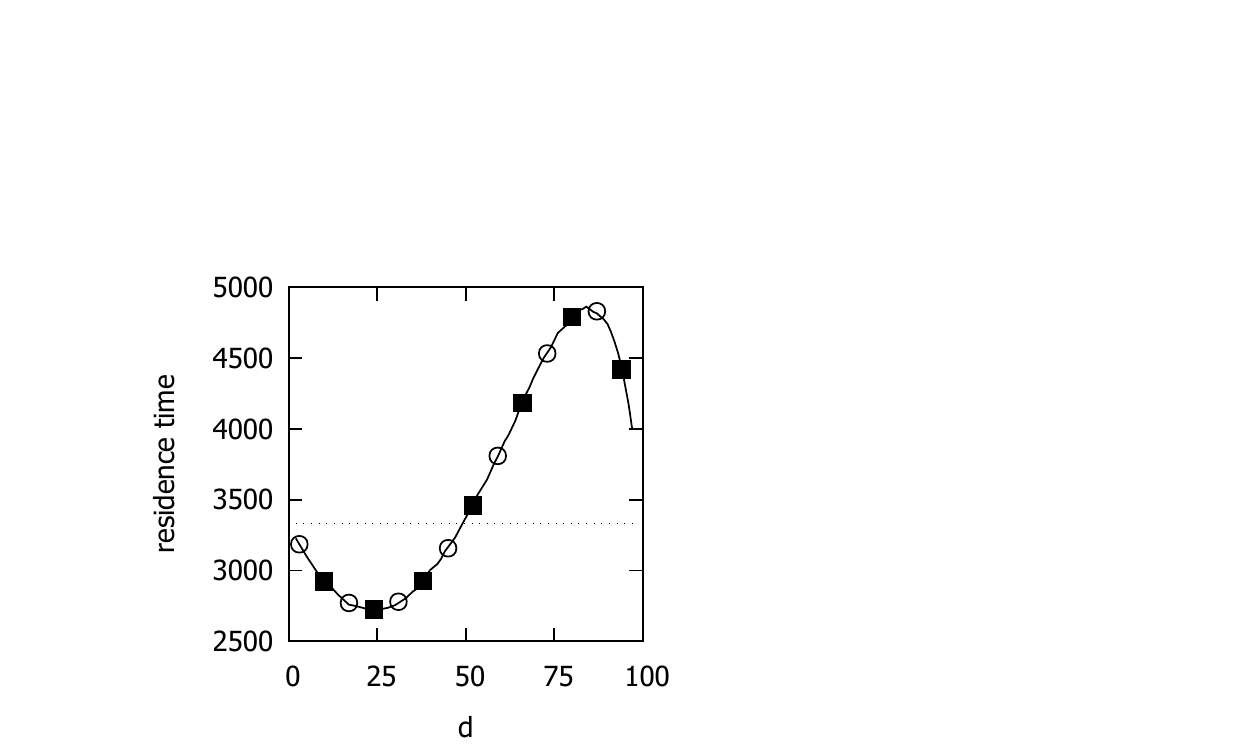}
}
\put(-50,-20){
  \includegraphics[width=0.8\textwidth]{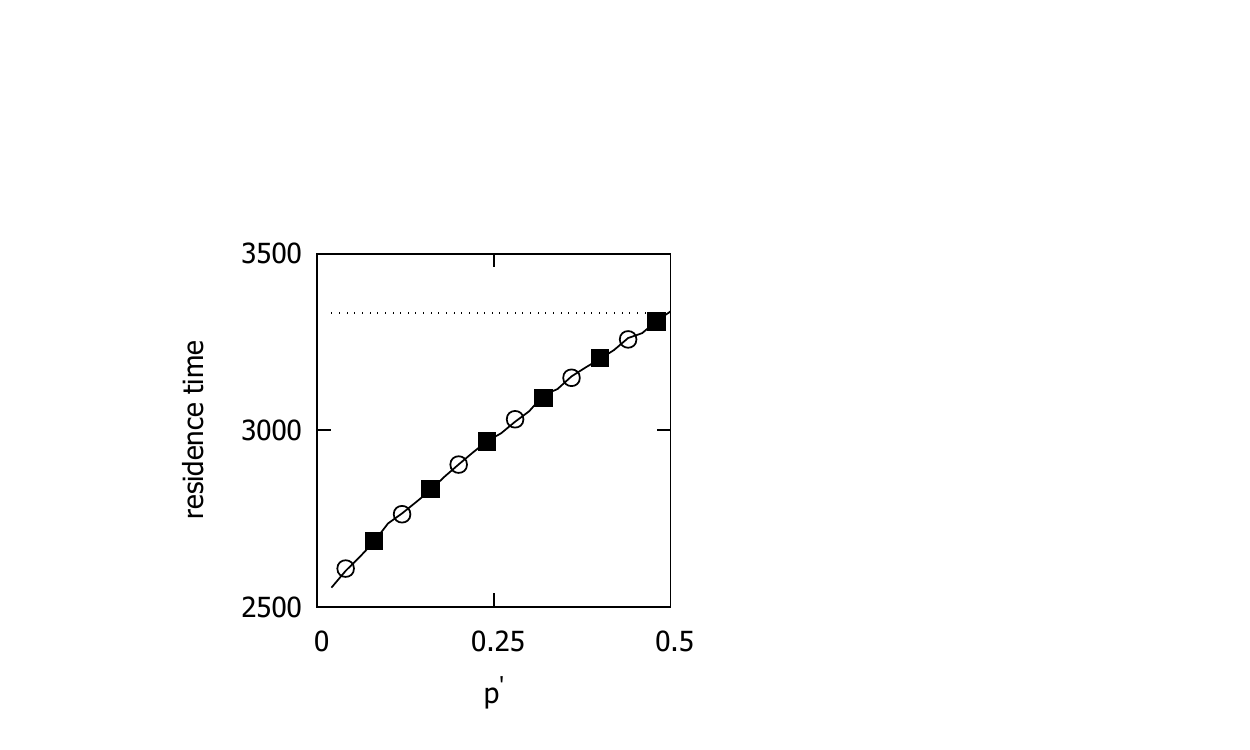}
}
\put(167,-20){
  \includegraphics[width=0.8\textwidth]{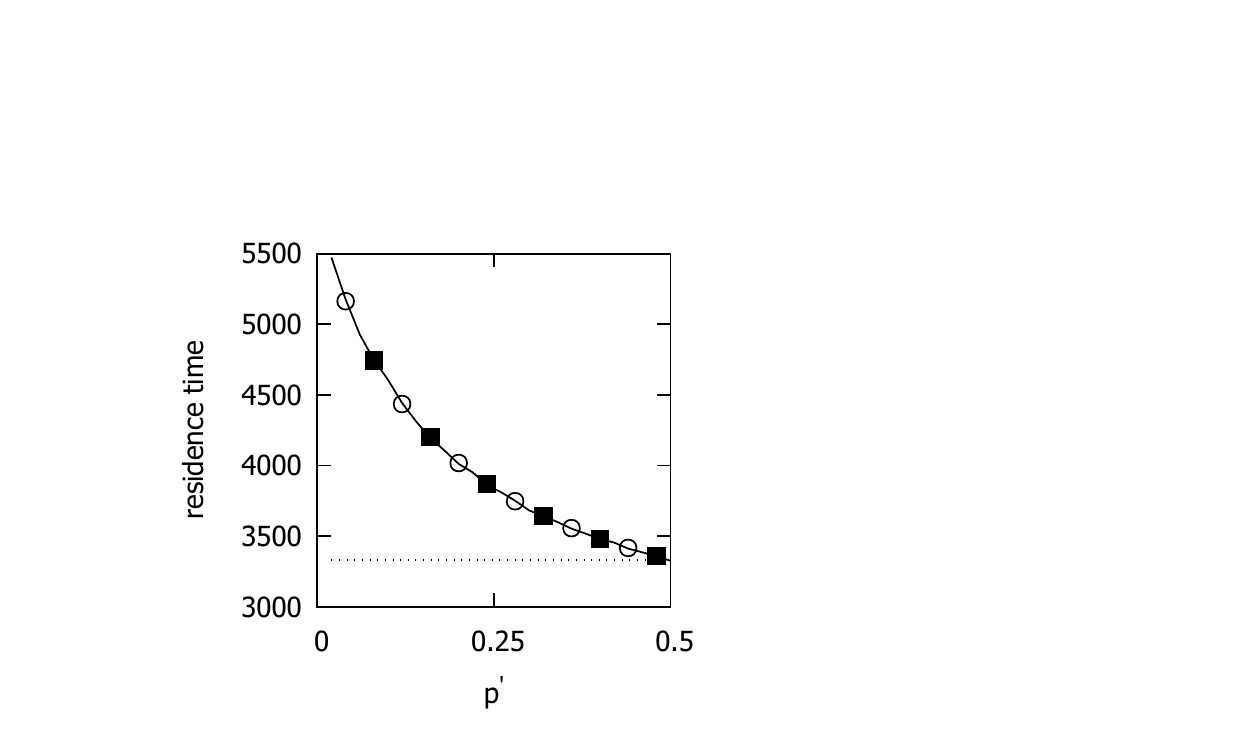}
}
\end{picture}
\caption{Residence time for $L=100$, $p=q=1/2$, and $q'=1/2$.
the residence time is computed with the 
local times approach for $p'\in[0.005,0.5]$ and $d$ ranging from $2$ to $98$.
In the other pictures the solid line represents the Monte Carlo measure
obtained simulating $10^7$ particles, whereas solid (resp.\ open) 
dots represent the generating function (resp.\ local time method) 
exact computation. The dashed line denotes the value 3333 of 
the residence time measured in absence of defect (this value is in 
yellow in the top left picture).
In the top right picture $p'=0.1$. 
In the bottom pictures $d=24$ (left) and $d=74$ (right).
}
\label{f:001}
\end{figure}

We consider first regular symmetric sites, i.e., $p=q=1/2$,
with a defect mimicking a 2D obstacle placed in the right part of the channel, 
namely, when the particle is on the defect it jumps to the right with 
probability $p'\in (0,1/2]$, to the left with probability $1/2$, 
and does not move with probability $\varepsilon=1/2-p'$.
Figure~\ref{f:001} shows the effect that different values of $p'$ and 
$d$ have on residence time.
We present the plot of the residence time as a function of 
$p'\in(0,1/2]$ and $d\in\{2,\ldots,98\}$ in the left top panel, where
the residence time is calculated with the local time approach. 
We represent in yellow 
the value of the residence time measured in absence of defect.
The other pictures represent a comparison among the proposed methods 
of evaluation of the residence time. 
In the top right picture the values of $p'$, $q'$, 
and $\varepsilon$ are fixed respectively to $0.1$, $0.5$,
and $0.4$, while $d$ is varied.
In the bottom panels the residence time is plotted as a function of 
$p'$ with 
$d=24$ on the left and $d=74$ on the right.
The solid line represents the Monte Carlo results 
obtained simulating $10^7$ particles. The number of particles that eventually 
cross the lane and contribute to the calculation of the average residence 
time grows when $p'$ and $d$ increase and it varies in the simulations 
from about $2\times10^4$ to $1\times10^5$.
The dashed line represents the value 3333 of the residence time 
in absence of defect. 

In the case considered in Figure~\ref{f:001} the defect is difficult 
to be overtaken by the particle since $p'\le1/2$, thus in the typical 
trajectories the walker 
spends a lot of time in the region before the defect. 
If the defect is close to the left end, it is probable that the particle 
finally exits the lane through the left side and it does not contribute to the 
residence time.
Thus, in this case the defect selects those particles that quickly 
overtake it and do not visit again the left part of the lane.
This explains why the residence time, for small $d$ and $p'$, is smaller than 
the no defect value. 
When $d$ is large, namely, the defect is close to the right end of the 
lane, it becomes more likely that a particle wanders a long time 
in the left part of the lane before 
eventually reaching the right side exit. 
This explains why the residence time for $d>L/2$ is larger than the 
no defect value. These effects are all negligible when $p'$ 
approaches $1/2$ and so the case of absence of defect is recovered.

\begin{figure}[ht!]
\begin{picture}(120,140)(0,0)
\put(60,-12){
  \includegraphics[width=0.8\textwidth]{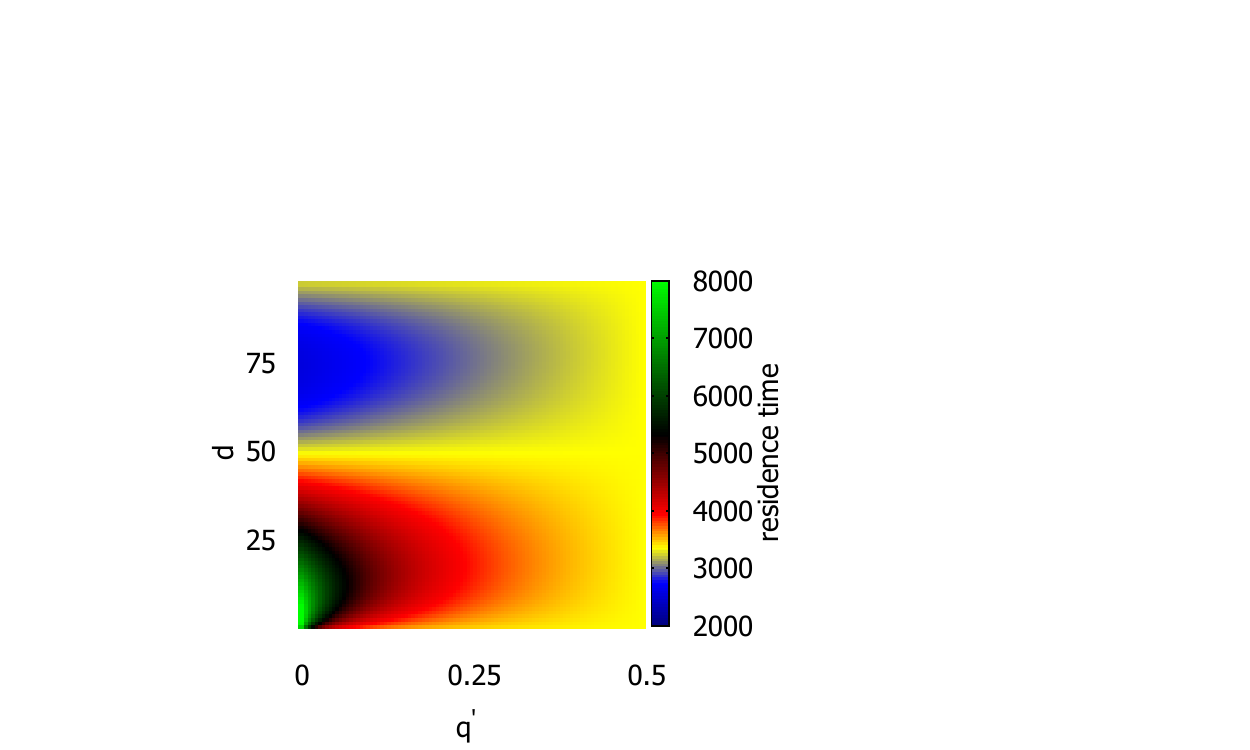}
}
\end{picture}
\caption{Residence time computed with the local times approach for
$L=100$, $p=q=1/2$, $q'\in[0.005,0.5]$, $p'=1/2$, $\varepsilon=0.5-q'$, 
$d=2,\dots,98$. 
The residence time value 3333 for the no defect case is represented 
in yellow.
}
\label{f:002}
\end{figure}

In Figure~\ref{f:002} we consider a case which is specular with respect 
to the one considered in Figure~\ref{f:001}:
we set $p'=1/2$ and vary $q'\in (0,0.5]$ with $\varepsilon=1/2-q'$.
Even in this case different values of $d\in\{2,\ldots,98\}$ are considered.
Obviously the results that we find are specular with respect to the 
line $d=L/2$ with respect to those depicted in the left top panel 
of Figure~\ref{f:001}.

\begin{figure}[ht!]
\begin{picture}(200,140)(0,0)
\put(-65,-12){
  \includegraphics[width=0.75\textwidth]{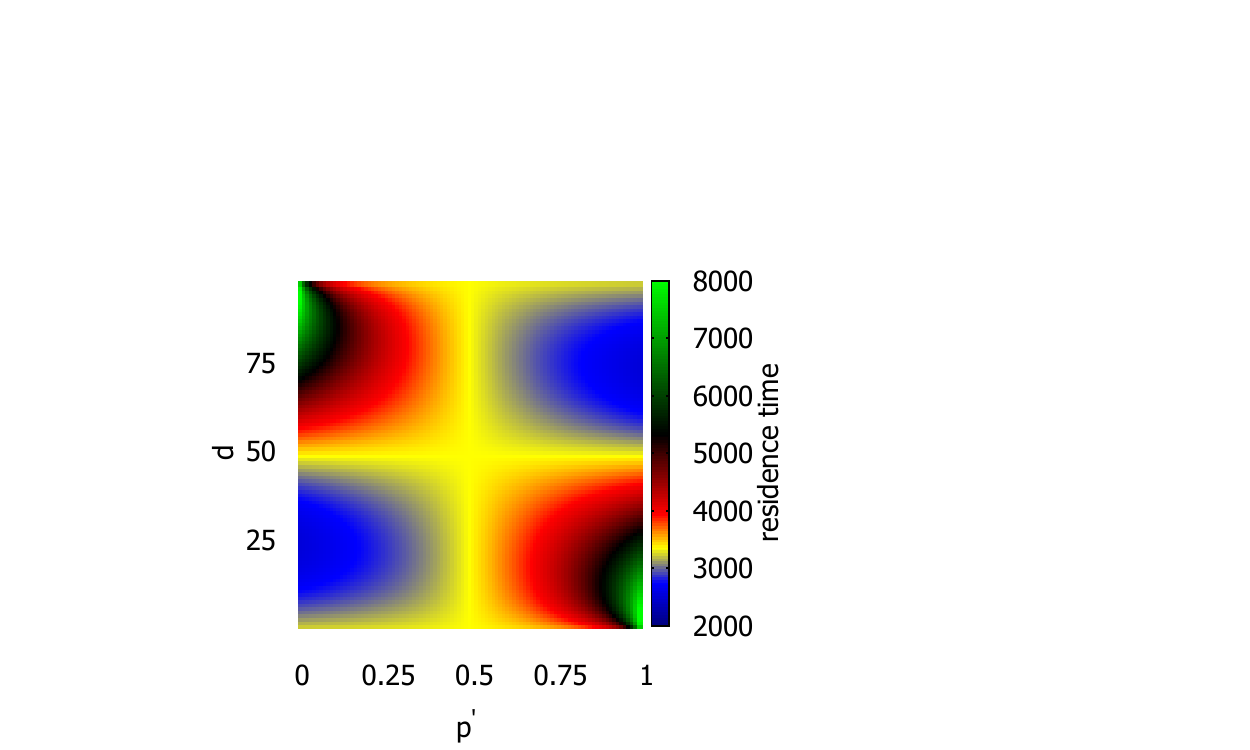}
}
\put(120,-10){
  \includegraphics[width=0.75\textwidth]{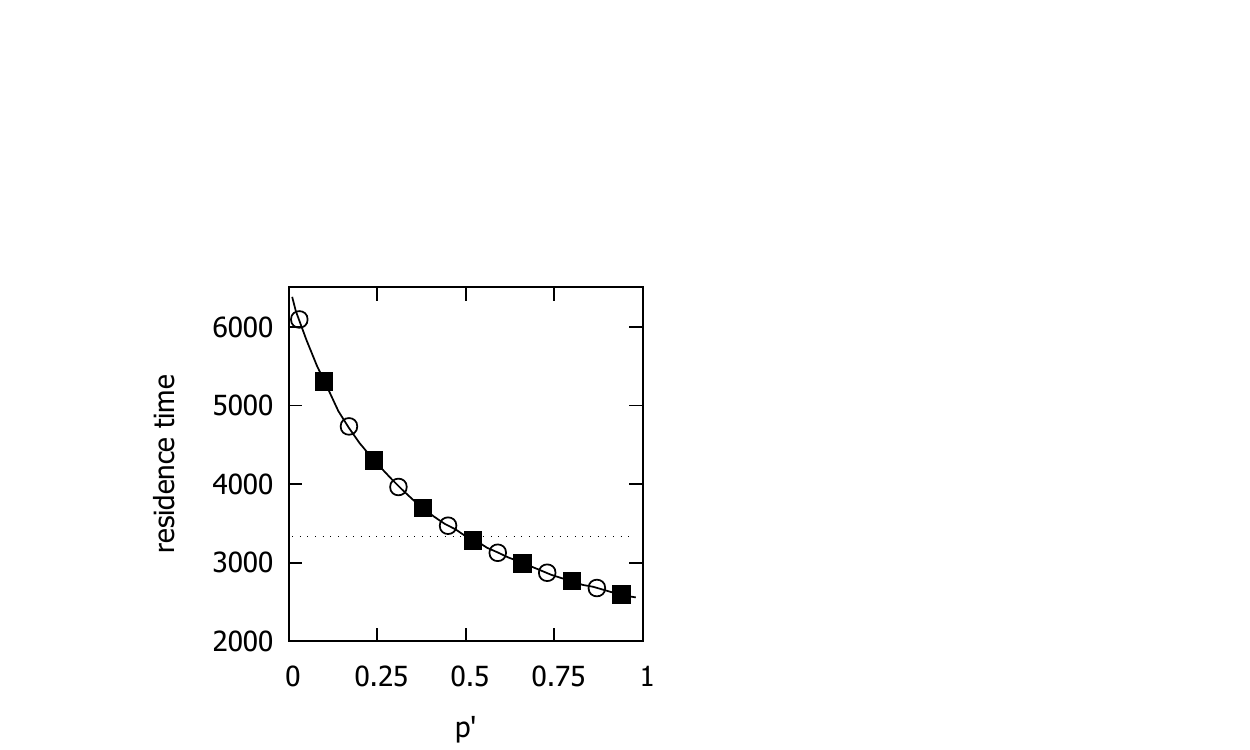}
}
\put(276,-10){
  \includegraphics[width=0.75\textwidth]{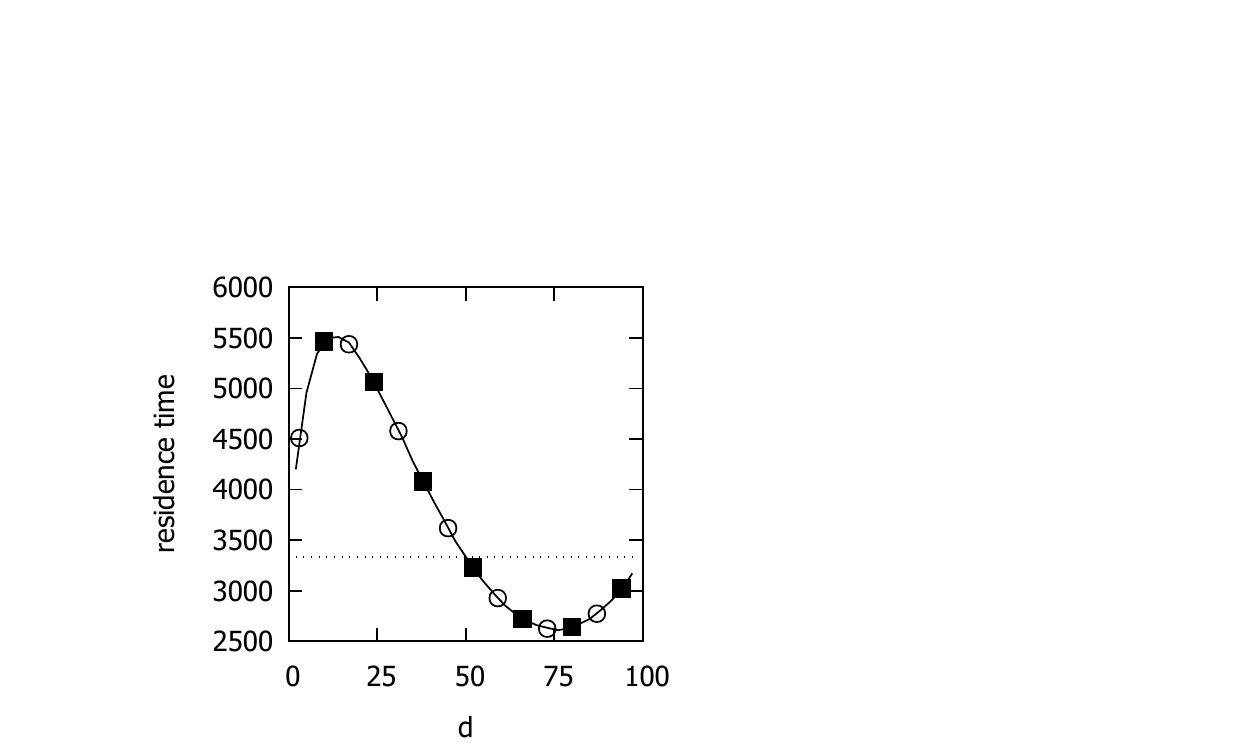}
}
\end{picture}
\caption{Residence time for $L=100$, $p=q=1/2$, $\varepsilon=0$. 
On the left the residence time is computed via the local times approach
for $p'\in[0.01,0.99]$ and $d=2,\dots,98$. The no defect value 3333
of the residence time is represented in yellow.
Center panel: residence time vs.\ $p'$  for $d=80$.
Right panel: residence time vs.\ $d$  for $p'=0.9$.  
For the center and right panel 
lines and symbols are as in Figure~\ref{f:001}.
}
\label{f:003}
\end{figure}

Still for symmetric regular sites, i.e., $p=q=1/2$, 
in Figure~\ref{f:003} we consider $p'=1-q'\in(0,1)$, so that 
the probability that the particle at the defect does not move is 
$\varepsilon=0$.
The behavior is coherent with that one discussed in the previous cases, 
with shorter or larger residence time depending on the position of the 
defect and the ratio $p'/q'$. If $p'=q'=1/2$ the no defect case is 
obviously recovered.
Note, and this is less intuitive, that the residence time is the same of the totally symmetric case 
 also if $d=50$, namely, when the defect is at the 
middle of the lane. 
As we did before, a comparison among results obtained via 
different approaches is proposed in Figure~\ref{f:003}.

\begin{figure}[ht!]
\begin{picture}(200,290)(-30,0)
\put(-50,142){
  \includegraphics[width=0.8\textwidth]{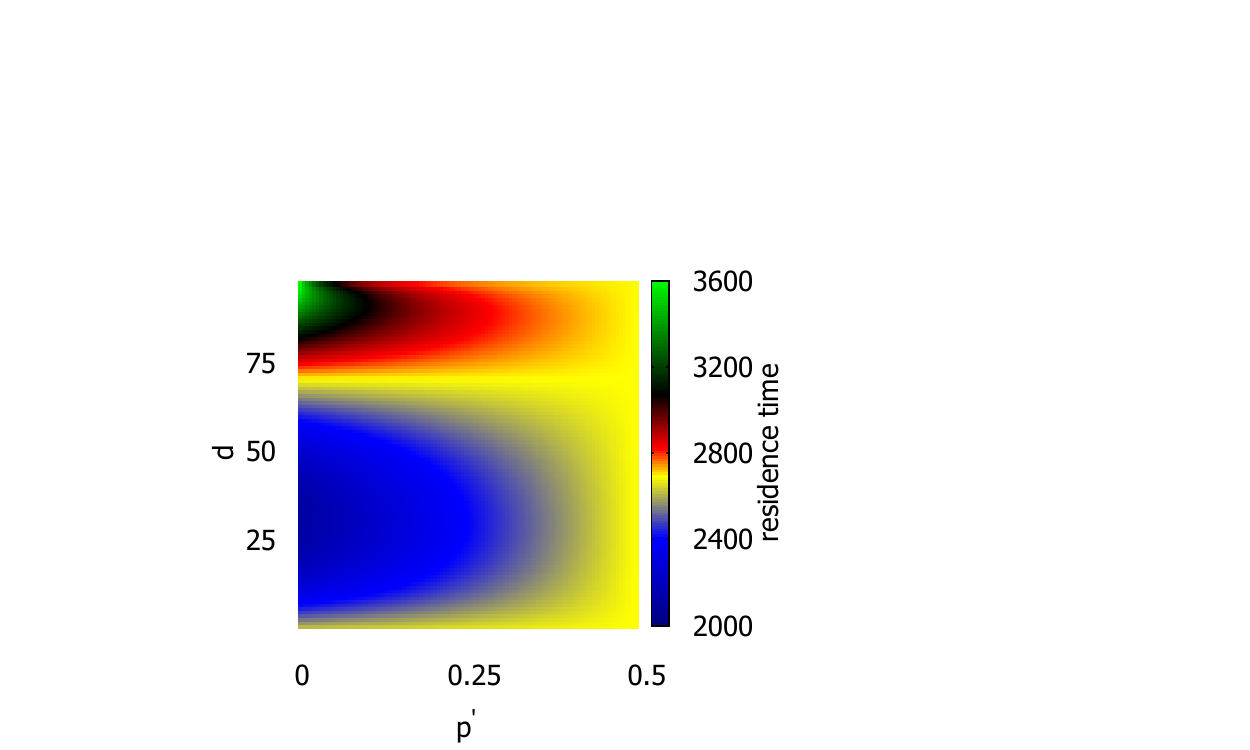}
}
\put(175,142){
  \includegraphics[width=0.8\textwidth]{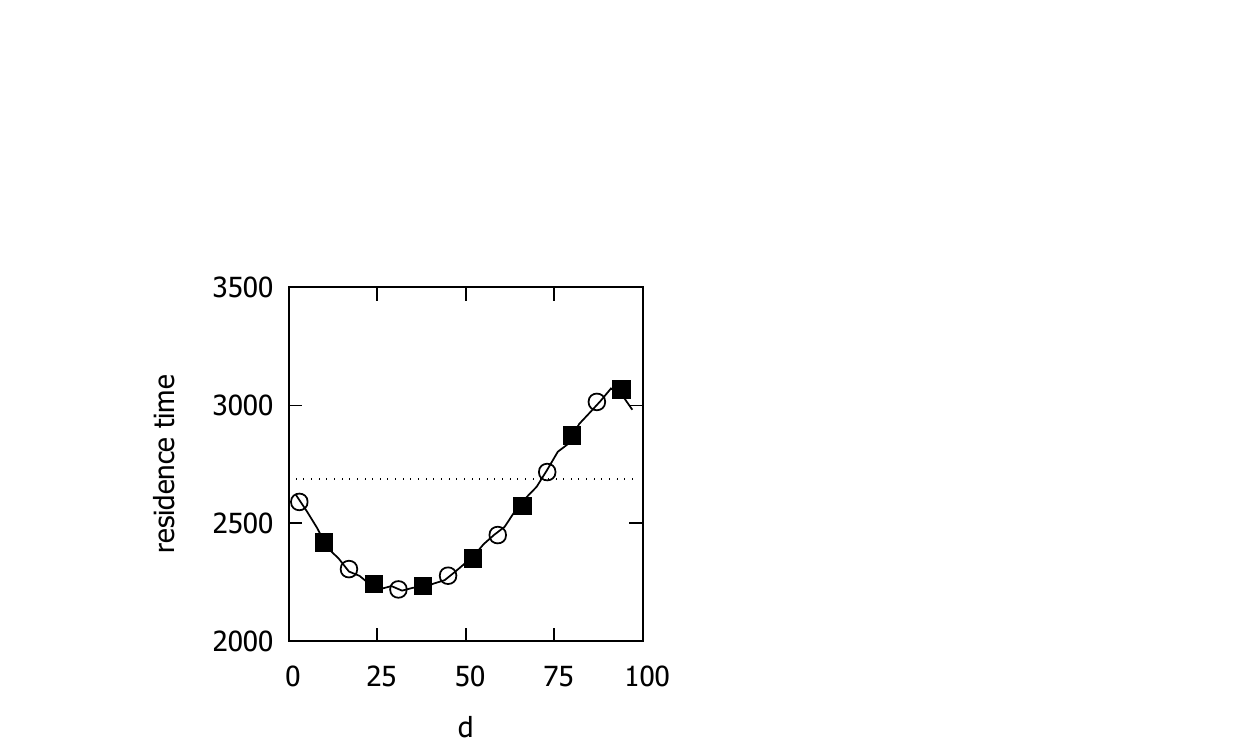}
}
\put(-50,-20){
  \includegraphics[width=0.8\textwidth]{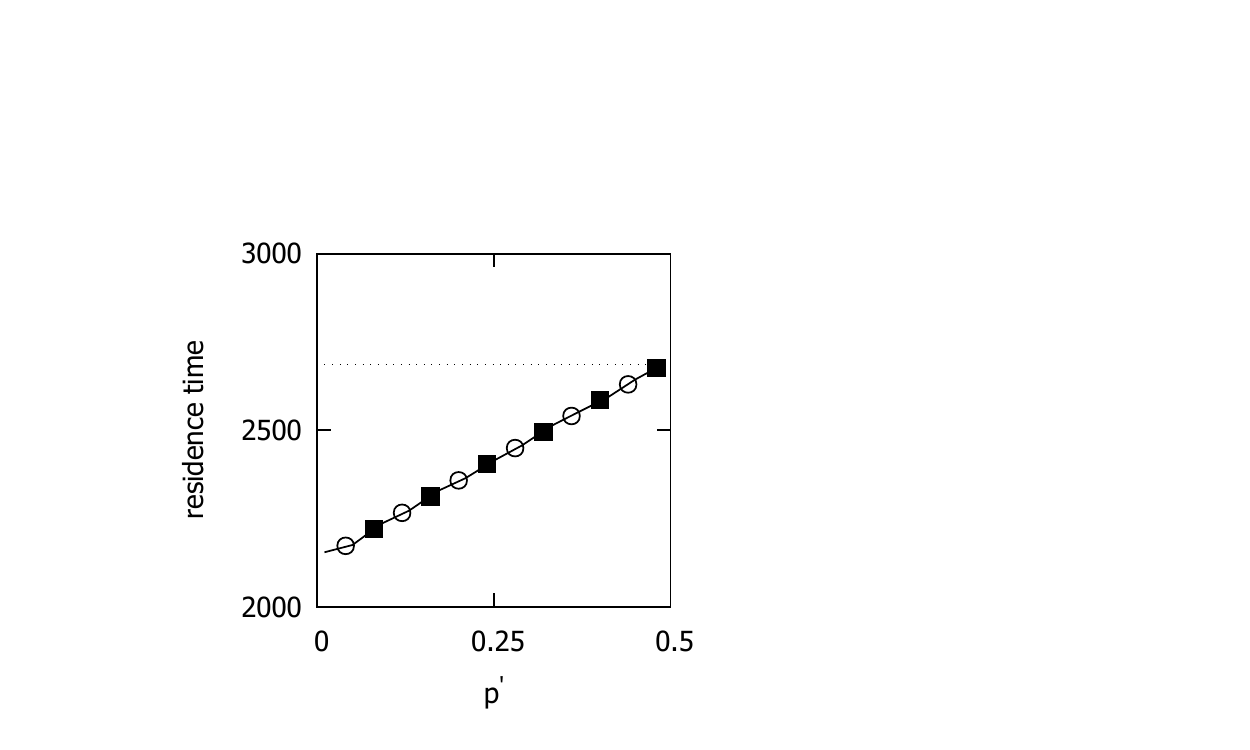}
}
\put(167,-20){
  \includegraphics[width=0.8\textwidth]{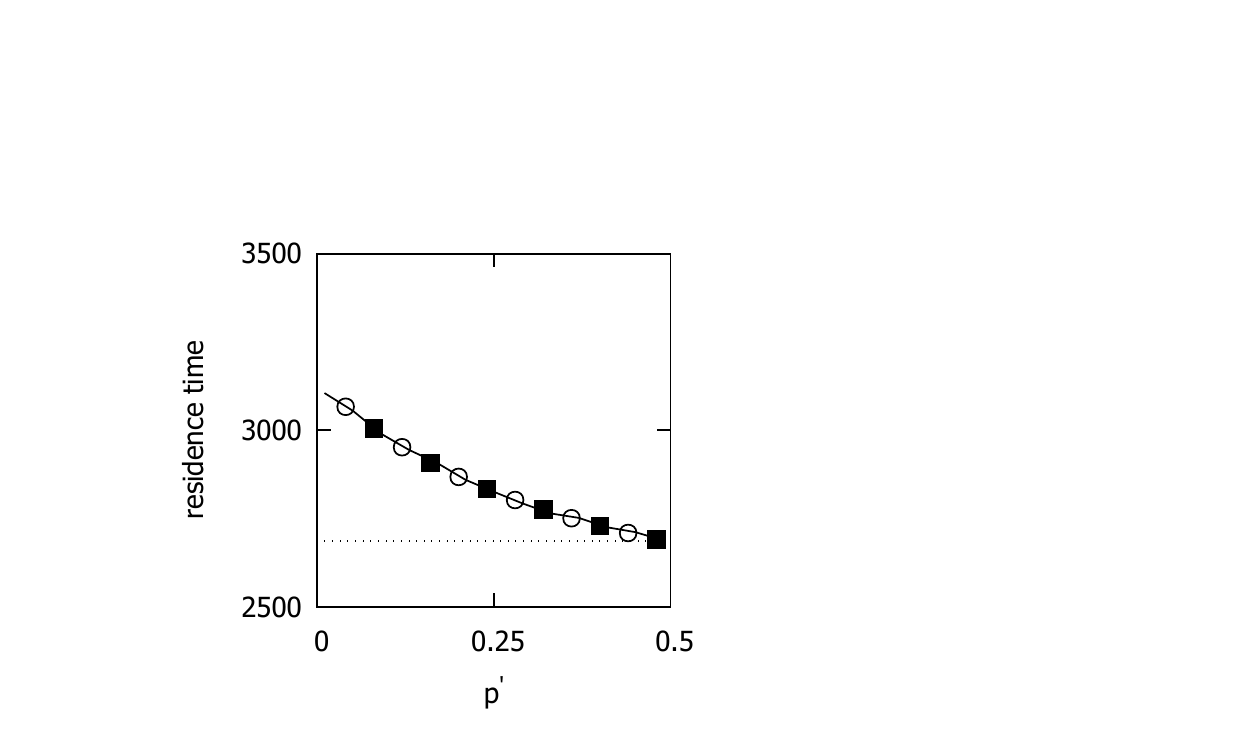}
}
\end{picture}
\caption{Residence time for $L=100$, $p=0.49$, $q=0.51$, $q'=q=0.51$.
The top left panel represents the residence time computed with the 
local times approach for $p'\in[0.005,0.49]$ and 
$d=2,\dots,98$.
Lines and symbols are as in Figure~\ref{f:001}, but the residence time 
with no defect is now 2686 and the total number of simulated 
particles is $10^8$.
In the top right panel $p'=0.1$.
In the bottom panels 
$d=24$ (left) and $d=85$ (right).
}
\label{f:004}
\end{figure}

The effects shown in Figure~\ref{f:003} for $p'\leq 1/2$ and 
$p'\geq 1/2$ are qualitatively the same as those in Figures~\ref{f:001} 
and \ref{f:002}, respectively.
Indeed, since $p'\leq 1/2$ implies $q'\geq p'$, in this case 
it is more difficult to overtake the defect than to be rejected.
Thus, the explanation of the residence time behavior 
is the same as for the case in Figure~\ref{f:001}. 
Conversely, if $p'\geq 1/2$ the behavior is similar to that in 
Figure~\ref{f:002}:
for a particle placed on the defect site it is more likely to jump to 
the right part of the lane and 
this favors a quick evacuation of the walker if the defect is beyond the 
middle point of the lane, while it makes difficult for a particle to 
exit easily from the left side if the walker meets the defect for $d<L/2$.

We now investigate the case in which the regular sites are not symmetric 
anymore, i.e., $p\neq q$. 
We first consider 
the analog of the case in Figure~\ref{f:001}, namely, we fix  
$p$ and $q=1-p$ and vary the defect parameters. 
At the defect the walker
jumps to the 
left as at regular sites, i.e., $q'=q$, but it jumps to the right 
with a different probability
$p'\in (0,p]$; as usual $\varepsilon=1-q'-p'$.
In Figure~\ref{f:004} we pick $p=0.49$ and $q=0.51$ and construct the plots 
analogous to those in Figure~\ref{f:001}.
The behavior is qualitatively similar to the one in Figure \ref{f:001}: 
for small values of $p'$ and $d$ the residence time is smaller with respect 
to the no defect case, but at small $p'$ it increases when $d$ is increased.
The presence of a small adverse drift, i.e., $p<q$, 
increases the value of $d$ at which the transition between the 
two effects is observed.
As remarked above, 
the plots in Figure~\ref{f:004} are similar to those in Figure~\ref{f:001},
but to perform the Monte Carlo computation many more 
particles had to be simulated
to average on a similar number of particles crossing the lane, 
 with respect to the cases considered before 
with symmetric regular sites, due to the 
sensible decreasing of the probability to cross the lane when $p<q$.

\begin{figure}[ht!]
\begin{picture}(200,290)(0,0)
\put(-50,130){
  \includegraphics[width=0.8\textwidth]{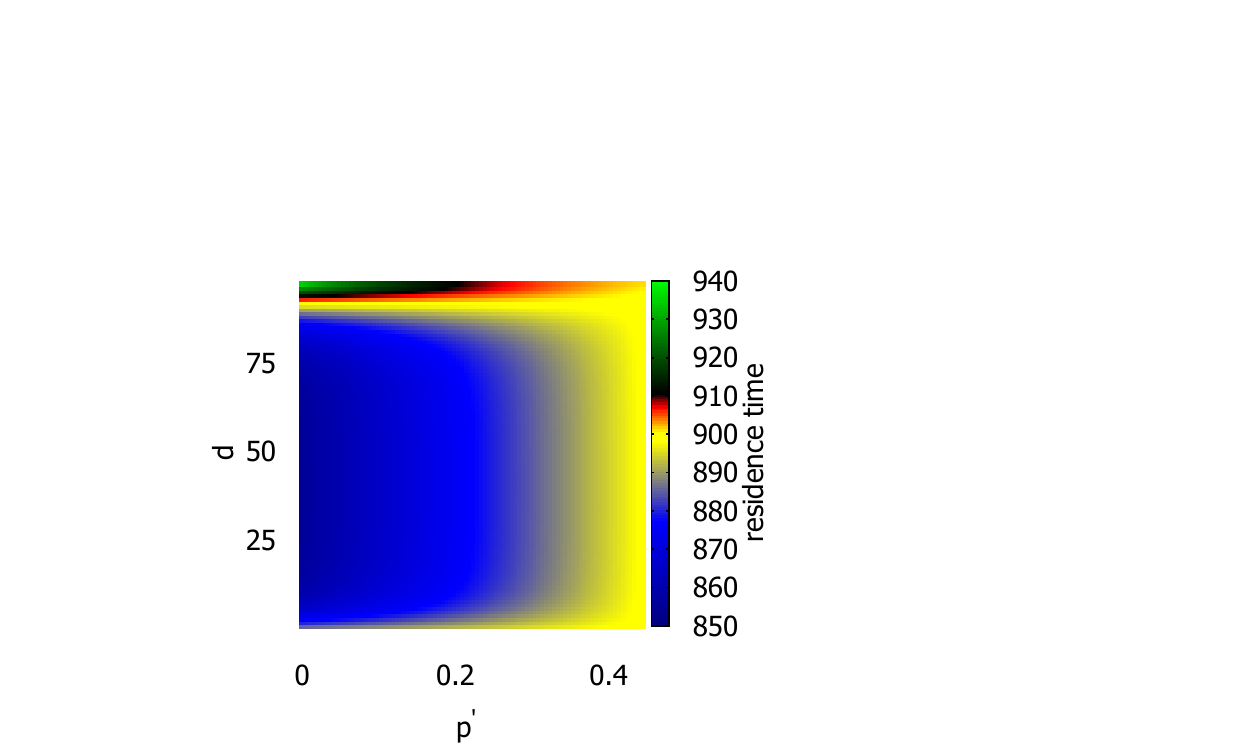}
}
\put(175,130){
  \includegraphics[width=0.8\textwidth]{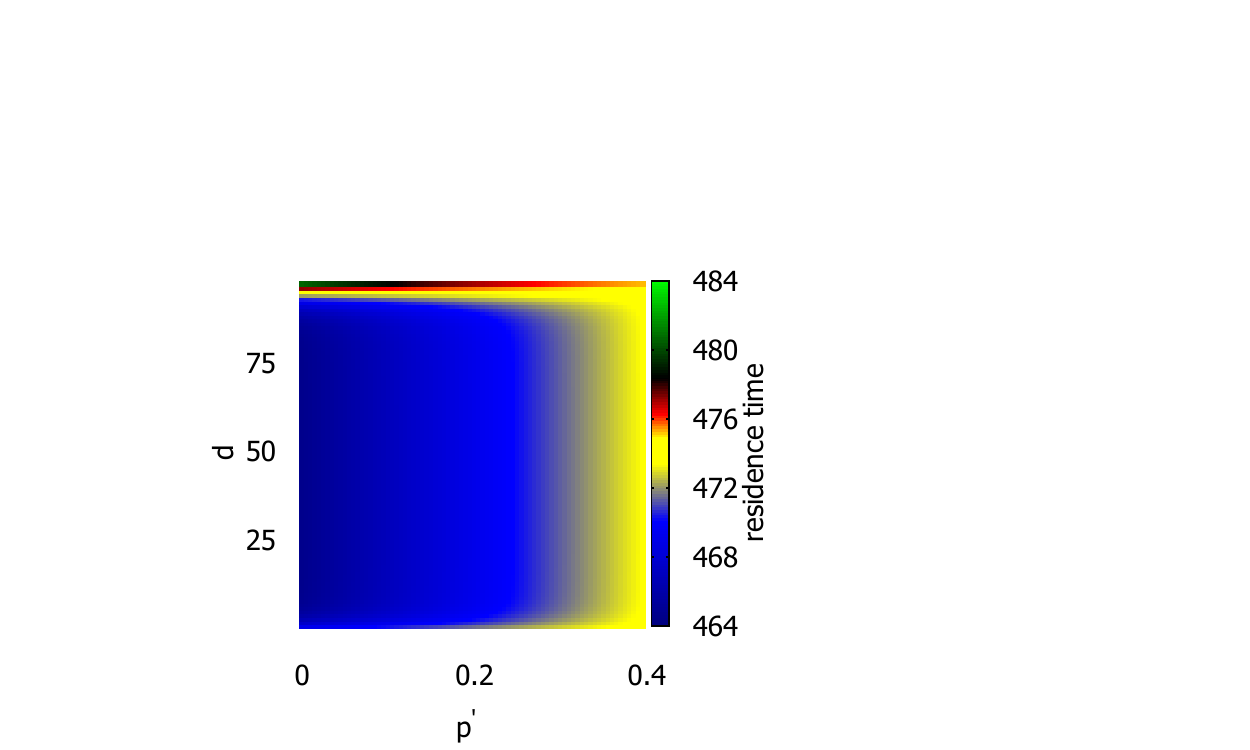}
}
\put(-50,-20){
  \includegraphics[width=0.8\textwidth]{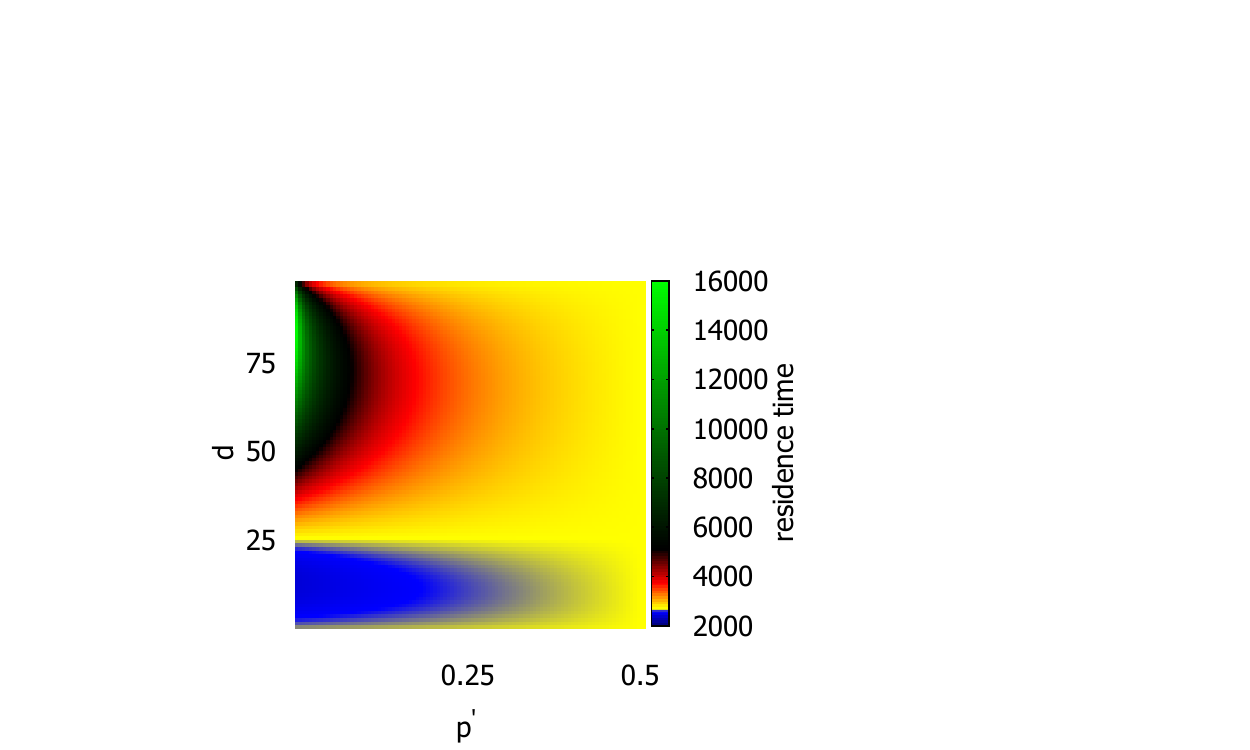}
}
\put(175,-20){
  \includegraphics[width=0.8\textwidth]{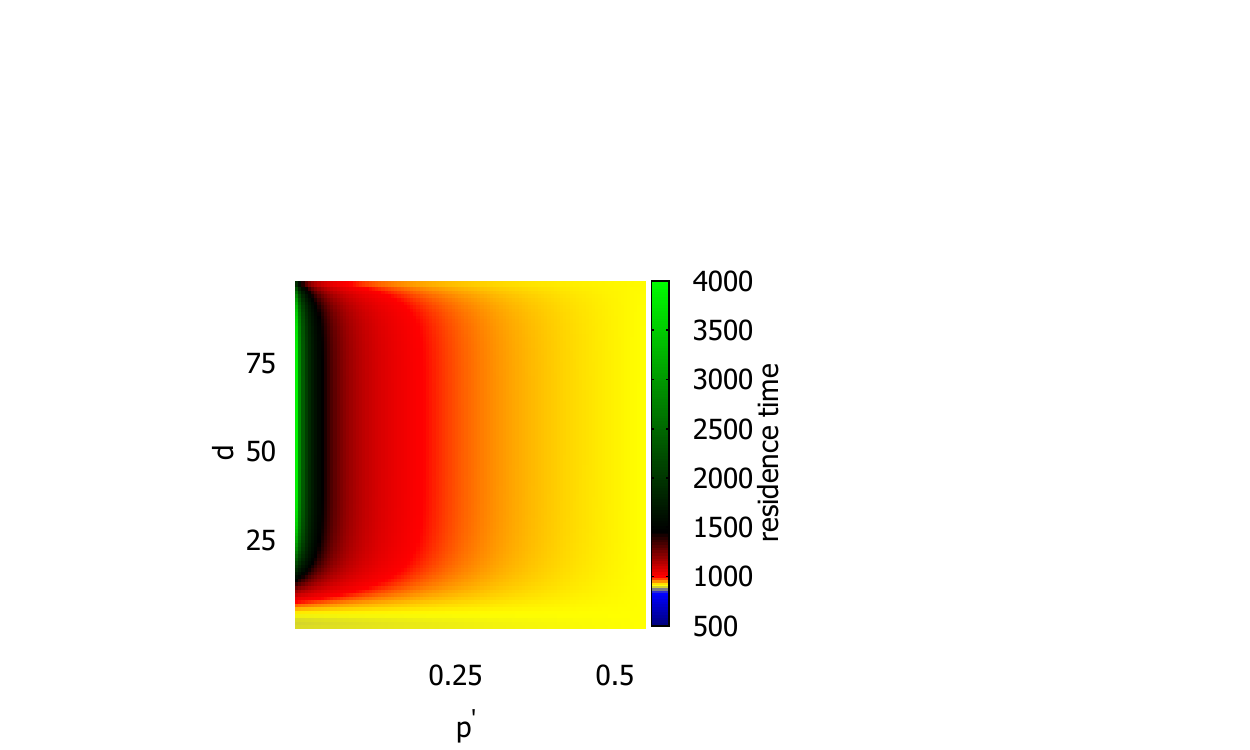}
}
\end{picture}
\caption{Residence time computed with the local times approach for 
$L=100$ and  $d=2,\dots,98$.
Top left panel: $p=0.45$, $q=0.55$, $p'\in[0.005,0.45]$, $q'=q$; the 
value of the residence time in absence of defect is $900$.
Top right panel: $p=0.4$, $q=0.6$, $p'\in[0.005,0.4]$, $q'=q$; the 
value of the residence time in absence of defect is $475$.
Bottom left panel: $p=0.51$, $q=0.49$, $p'\in[0.005,0.51]$, $q'=q$; the 
value of the residence time in absence of defect is  $2686$.
Bottom right panel: $p=0.55$, $q=0.45$, $p'\in[0.005,0.55]$, $q'=q$; the 
value of the residence time in absence of defect is $900$.
In yellow it is represented the residence time with no defect ($p'=p$).
}
\label{f:005}
\end{figure}

\begin{figure}[ht!]
\begin{picture}(200,140)(0,0)
\put(-50,-10){
  \includegraphics[width=0.8\textwidth]{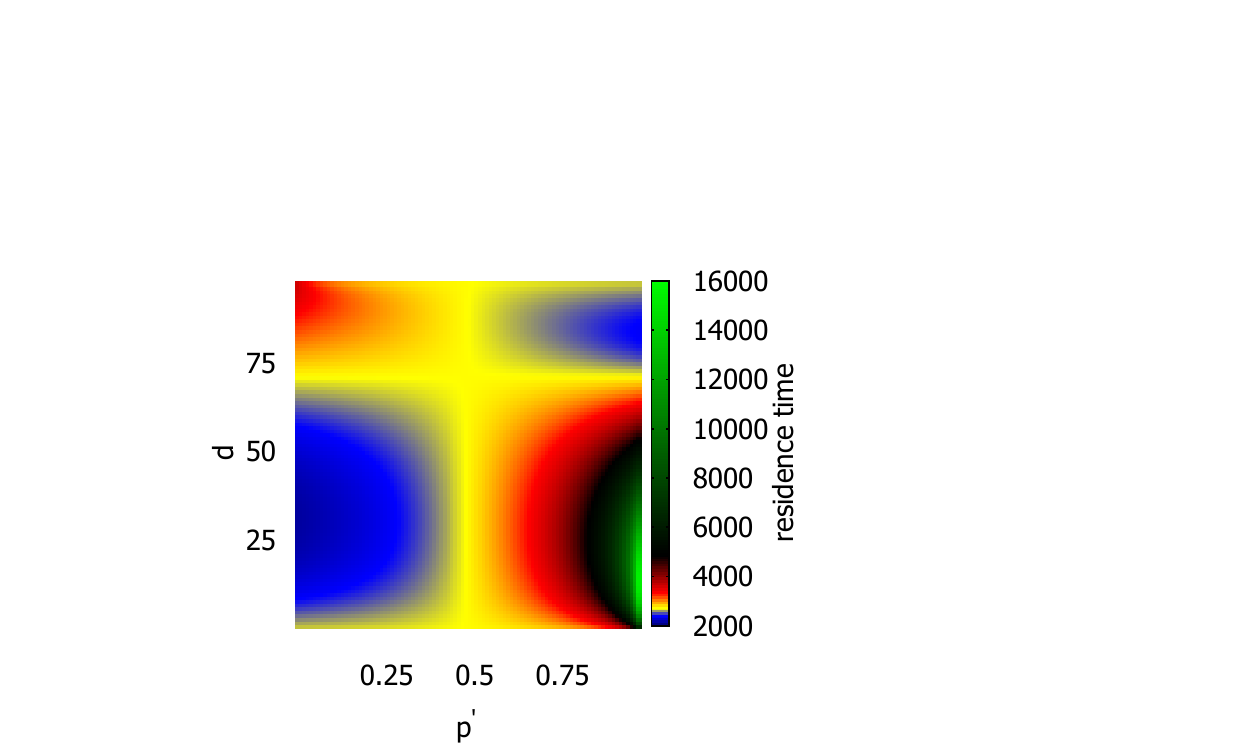}
}
\put(175,-10){
  \includegraphics[width=0.8\textwidth]{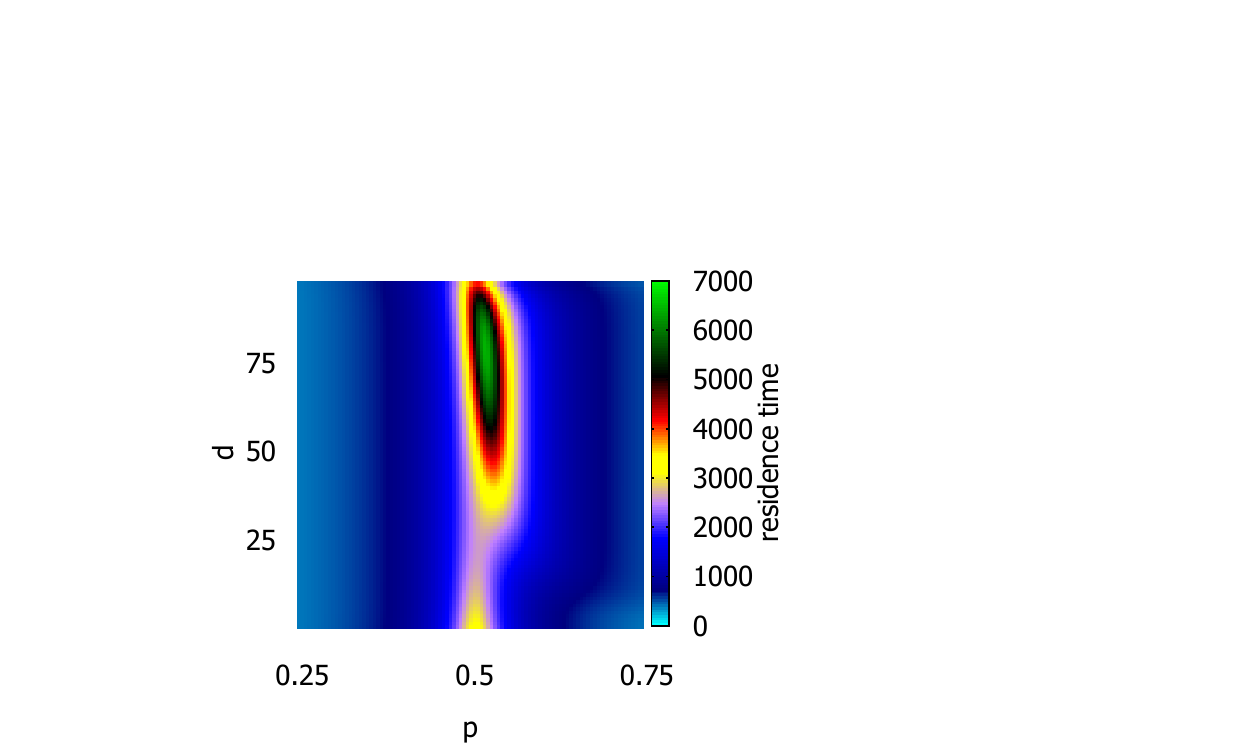}
}
\end{picture}
\caption{Residence time computed with the local times approach 
for $L=100$ and $d=2,\dots,98$.
Left  picture: $p=0.49$, $q=0.51$, $p'\in[0.01,0.99]$, $q'=1-p'$, the 
value of the residence time in absence of defect is  $2686$ (in yellow).
Right  picture:  $p'=0.1$, $q'=0.9$, $p\in[0.25,0.75]$, $q=1-p$.
}
\label{f:ultima}
\end{figure}

Different values of the drift at regular sites modify further the 
residence time behavior. If $p$ decreases only values of $d$ very close 
to $L$ produce a larger residence time (see Figure \ref{f:005}, top row, 
where $p$ is equal to $0.45$ and $0.4$ respectively). 
If $p$ is larger than $1/2$, it is possible to observe a smaller 
residence time only if the drift is small, for small values of $d$. 
In the pictures in Figure \ref{f:005} bottom row, for $p=0.51$ we can 
still see particles crossing the lane faster for small $d$ and $p'$, 
while for $p=0.55$ this effect is negligible.
In each of these plots it is depicted in yellow
the value of residence time for $p'=p$ (no defect case).
Note that the presence of drift, both if $p<q$ or not, 
produces in absence of defect a notable reduction of the residence time,
 see also the right panel in Figure \ref{f:ultima}.

As we did in Figure~\ref{f:002} for the regular symmetric sites case, 
we could consider the analogous case even in presence of drift. Since the 
effect is similar we do not report the data in details. 

Finally, we summarize our results in Figure~\ref{f:ultima}, where  left and right panels report, 
respectively, the dependence of the residence time on $p'$ and $p$ for any value of $d$. 
In Figure~\ref{f:ultima} left panel, 
where $\varepsilon=0$ (hence $p'=1-q'$),
we can notice a behavior similar to the one in Figure~\ref{f:003},
 but  the drift modify the value of $d$ for which the transition 
 between the zones with larger and smaller residence time happens, 
 as observed in Figures~\ref{f:004} and \ref{f:005}.
In the picture the small drift case $p=0.49$ and $q=0.51$  is considered.

The right panel, where a defect with $p'=0.1$ and $q'=0.9$ is considered, points out also
that the presence of the drift yields smaller values of the 
residence time whatever the direction of the drift is. If the drift 
is directed 
towards the left end 
this is due to the fact that only fast particles are selected 
to exit through the right end of the lane. 
On the other hand, 
if the drift 
is directed 
towards the right end the effect is simply due to the fact that 
particles are pushed towards the right end of the lane.  

As a general remark we notice that the residence time is mainly 
influenced by the ratios between $q$ and $p$ and between $q'$ and $p'$.
The effect of the probability $\varepsilon$ to remain steady 
is to increase the residence time by letting grow the local times only 
on the defect site. This growth is in general not so consistent.
Similarly, the probability $r$ acts on regular sites and it produces a 
growth of a factor $1/(1-r)$ of the local times on each regular site.
So, the behaviors with different values of $r$ and $\varepsilon$ can 
be deduced from the cases presented above, and we will not show further 
plots for these cases.


%



\end{document}